\documentclass[fleqn,usenatbib]{mnras}

\usepackage{multirow}
\usepackage{colortbl}
\usepackage{color}
\usepackage[fleqn]{amsmath}
\usepackage{amssymb}
\usepackage{amsfonts} 
\usepackage{verbatim}
\usepackage{scalefnt}
\usepackage{lscape}
\usepackage{booktabs}
\usepackage{float}
\usepackage[normalem]{ulem}
\usepackage{natbib} 
\usepackage{adjustbox}
\usepackage{rotating}
\usepackage{gensymb}
\usepackage{soul}
\usepackage{graphicx}
\usepackage[dvipsnames]{xcolor}
\usepackage{comment}
\usepackage{siunitx}
\usepackage{scalerel,amssymb}
\usepackage{tikz}

\definecolor{palepink}{rgb}{0.98, 0.85, 0.87}
\definecolor{nadeshikopink}{rgb}{0.96, 0.68, 0.78}
\definecolor{lightpink}{rgb}{1.0, 0.71, 0.76}
\definecolor{babypink}{rgb}{0.96, 0.76, 0.76}
\definecolor{bubblegum}{rgb}{0.99, 0.76, 0.8}
\definecolor{cherryblossompink}{rgb}{1.0, 0.72, 0.77}
\definecolor{cottoncandy}{rgb}{1.0, 0.74, 0.85}
\definecolor{bleudefrance}{rgb}{0.19, 0.55, 0.91}

\definecolor{blue-violet}{rgb}{0.54, 0.17, 0.89}
\definecolor{darkgreen}{rgb}{0,0.6,0}
\definecolor{ballblue}{rgb}{0.13, 0.67, 0.8}
\definecolor{gray}{rgb}{0.4,0.5,0.6}
\definecolor{green(html/cssgreen)}{rgb}{0.0, 0.5, 0.0}

\usepackage{hyperref}
\newcommand{\beq}{\begin{equation}}
\newcommand{\eeq}{\end{equation}}
\usepackage{amsmath} 

\usepackage{longtable}

\title[JWST imaging of dust beyond the torus]{
Dust beyond the torus: Revealing the mid-infrared heart of
local Seyfert ESO 428-G14 with JWST/MIRI
}
\author[Haidar et al. ]{
Houda Haidar,$^{1}$ \thanks{E-mail: h.haidar2@newcastle.ac.uk}
David J. Rosario,$^{1}$ 
Almudena Alonso-Herrero,$^{2}$
Miguel Pereira-Santaella,$^{3}$
\newauthor 
Ismael Garc\'ia-Bernete,$^{4}$ 
Stephanie Campbell,$^{1}$ 
Sebastian F. H\"onig,$^{5}$ 
Cristina Ramos Almeida,$^{6,7}$
\newauthor 
Erin Hicks,$^{8,9,10}$ 
Daniel Delaney,$^{8,9}$ 
Richard Davies,$^{11}$ 
Claudio Ricci,$^{12,13}$ 
Chris M. Harrison,$^{1}$ 
\newauthor 
Mason Leist,$^{10}$
Enrique Lopez-Rodriguez,$^{14}$ 
Santiago Garcia-Burillo, $^{15}$
Lulu Zhang,$^{10}$
Chris Packham,$^{10, 16}$
\newauthor 
Poshak Gandhi,$^{5}$ 
Anelise Audibert,$^{6,7}$
Enrica Bellocchi,$^{17,18}$
Peter Boorman,$^{19,20}$ 
Andrew Bunker,$^{4}$
\newauthor
Françoise Combes,$^{21}$ 
Tanio Diaz Santos,$^{22,23}$
Fergus R. Donnan,$^{4}$
Omaira Gonzalez Martin,$^{24}$
\newauthor
Laura Hermosa Mu\~noz,$^{2}$
Matthaios Charidis,$^{1}$
Alvaro Labiano,$^{25}$
Nancy A. Levenson,$^{26}$
Daniel May,$^{27}$
\newauthor
Dimitra Rigopoulou,$^{4,28}$
Alberto Rodriguez Ardila,$^{29,30}$
T. Taro Shimizu,$^{11}$
Marko Stalevski,$^{31,32}$
\newauthor
Martin Ward$^{33}$  \\
\hfill \\
\noindent (Affiliations are listed at the end of the paper)}

\date{Accepted XXX. Received YYY; in original form ZZZ}

\pubyear{2024}

\begin{document}
\label{firstpage}

\maketitle

\begin{abstract}
Polar dust has been discovered in a number of local Active Galactic Nuclei (AGN), with radiation-driven torus models
predicting a wind to be its main driver. However, little is known about its characteristics, spatial extent, or connection
to the larger scale outflows. We present the first JWST/MIRI study aimed at imaging polar dust by zooming onto the centre of ESO 428-G14, part of the  Galaxy Activity, Torus, and Outflow Survey (GATOS)  survey of local AGN. We detect extended mid-infrared (MIR) emission within 200 pc from the nucleus. This polar structure is co-linear with a radio jet and lies perpendicular to a molecular gas lane that feeds and obscures the nucleus. Its morphology bears a striking
resemblance to that of gas ionised by the AGN in the narrow-line region. We demonstrate that part of this spatial correspondence is due to contamination within the JWST filter bands from strong emission lines. Correcting for the contamination, we find the morphology of the dust continuum to be more compact, though still clearly extended out to $r\approx 100 \, \rm pc$. We estimate the emitting dust has a temperature of $\sim 120\,\rm K$. Using simple models, we find that the heating of small dust grains  by the radiation from the central AGN and/or radiative jet-induced shocks is responsible for the extended MIR emission.  Radiation-driven dusty winds from the torus is unlikely to be important. This has important implications for scales to which AGN winds can carry dust and dense gas out into their host galaxies.

\end{abstract}

\noindent

\section{introduction}

For decades, the dusty torus  has been held responsible for the dichotomy between type\,1 and type\,2  Active Galactic Nuclei (AGN), forming the  keystone of AGN unification \citep{Antonucci+85,Urry+95}.
Therefore, our insights into the AGN model and unification theory owe much to our ability to constrain the torus, and our understanding of dusty material in the vicinity of supermassive black holes (SMBHs). Pioneering observations with the Atacama Large Millimeter/submillimeter Array (ALMA) have, for the first time, enabled the detection and resolution of parsec-scale dusty molecular tori ($\sim 3$ to $50\, \rm pc$)  first in NGC1068 and later on in some local AGN \citep{GarciaBurillo+16,GarciaBurillo+19,GarciaBurillo+21,Gallimore+16,Alonso-Herrero+18,Alonso-Herrero+19,Alonso-Herrero+23,Combes+19}.
At the same time, images of the torus are now being produced by the new generation of  near- and mid-infrared (MIR) interferometers \citep{GRAVITY+20,Gamez+22}.\\

The AGN-heated dust in the torus spans a range of temperatures from
tens to  $ \sim 1800\, \rm K$, and peaks
in the MIR range  ($\lambda = 5-30\, \rm  \mu m$). This makes observations in the MIR part of the spectrum crucial for revealing the characteristics and structure of the torus. 
 MIR observations  have uncovered unprecedented details about the circumnuclear dust distribution in local Seyfert galaxies \citep[e.g.][]{Jaffe+04,Tristram2007,Burtscher+09,RamosAlmeida2009,RamosAlmeida2011,Reunanen2010,2014MNRAS.443.2766A,2016MNRAS.455..563A,GarciaBernete+15,GarciaBernete+16,GonzalezMartin+19b}. 
In some Seyfert galaxies, interferometric and  single-dish observations  revealed 
the dust to be distributed in a two-component system, with the majority of its MIR emission emerging from the polar direction 
\citep[e.g.][]{Honig+12,Honig+13,Tristram+14,LopezGonzaga2014,Leftly2018}.
Systematic studies have shown that the polar dust  can be traced from a few parsecs 
\citep[e.g.][]{LopezGonzaga+16,Leftly2018,Gamez+22,Isbell+22}
to as far as a few hundred parsecs from the central engine
\citep[e.g.][]{Radomski2003,packham2005,Asmus2014,Asmus+16,GarciaBernete+16,Alonso-Herrero2021}.  However, 
 the origin of the extended MIR emission (tens to hundreds of parsecs) is not clear yet, because it can include line emitting gas in the narrow line region (NLR), ambient dust in the NLR, and/or outflowing dust driven from the central region \citep{RamosAlmeida2017,Honig+19}. \\

Various torus models have been proposed to reproduce  the infrared emission of AGN and to help constrain torus properties and dynamics \citep[e.g.][]{Krolik+88,Emmering+92,Nenkova+08,Honig2010,GonzalezMartin+23}.
The detection of polar 
emission
motivated the inclusion of a polar component 
associated with dust driven by a wind, launched
from the inner parts of the torus, and powered by radiation pressure  
\citep[e.g.][]{Honig+12,Honig+13,Honig+17}.
For instance, radiative transfer models that incorporate  disk+wind components are able to reproduce the observed polar dust structures 
\citep[e.g.][]{Stalevski+17,Honig+17,Stalevski+19}.
In agreement with this, modern hydrodynamical simulations that incorporate radiative feedback predict polar dust to be a common feature in AGN, subject to high Eddington ratios 
\citep[e.g.][]{Wada+16,Williamson+20}.
The detection of a clear decrease in the fraction of obscured AGN at $\lambda_{\rm Edd} \geq 10^{-2}$  further confirms the idea that radiation pressure on dust is fundamental in shaping the environments of AGN \citep{Ricci+17}.
As such, a revised model of the obscuring torus has been put forward, 
where the torus is now described as a complex entity with multiple components and phases \citep[e.g.][]{RamosAlmeida2017,Izumi2018,Honig+19,Alonso-Herrero2021}. 
However, even if the dust is not part of a dusty wind, it can still substantially contribute to obscuration if it is just dust in the NLR  heated by the central engine \citep[e.g.][]{Radomski2003}.\\

Earlier ground-based work has reported that the silicate emission feature around $10\, \rm \mu m$ is absent in the polar dust \citep[e.g.][]{Honig+12,Honig+13,Burtscher+13}. This is best explained by the sublimation of small silicate grains and/or a dusty wind predominantly composed of large graphite grains \citep{Honig+12} which do not produce silicate features \citep[e.g.][]{1993ApJ...402..441L}.  A physical mechanism for the destruction of grains in a wind was proposed by \citet{2020ApJ...892..149T}. In this model, dust grains entrained in the wind are subject to hypersonic drift due to radiation pressure, which ultimately destroys them by kinetic sputtering. Smaller grains are more rapidly destroyed than larger grains \citep{1979ApJ...231...77D}, which results in only the larger grains surviving beyond the inner several parsecs \citep{2020ApJ...892..149T}. \\

While ground-based observations found compelling evidence of polar dust, they are still limited by observing conditions (e.g. high thermal background, unstable PSF) that make it difficult to resolve polar dust emissions,
and even if detected, the polar
structure can only be traced out to a few hundred parsecs \citep{Asmus+16}. In general, the challenge in detecting polar dust emissions lies not just in technical limitations but also in distinguishing these emissions from the surrounding galaxy, particularly for weaker AGN \citep{GonzalezMartin+15,GonzalezMartin+17}.
With the launch of the James Webb Space Telescope (JWST, \citealt{Gardner+23}), offering unparalleled resolution and sensitivity, we can now image and trace the extent of polar dust emission from  a parsec and a kiloparsec scale.  \\

With this aim, 
we present a multi-band imaging study of ESO 428-G14  with the MIRI instrument \citep{Rieke+15,Wright+15,Wright+23} covering a wavelength range from $5.6$ to $\rm 21\, \mu m$. ESO 428-G14 was selected as it shows prior evidence of polar MIR emission from ground-based observations \citep{Asmus+16}. 
 By combining our new MIRI data with the wealth and variety of ancillary photometry already available for ESO\,428-G14,
we are able to explore in great
detail the nature of its MIR emission, but also to place it within the broader context of this AGN.\\

\subsection{ESO 428-G14}
\begin{figure*}
    \centering
    \includegraphics[scale=0.6]{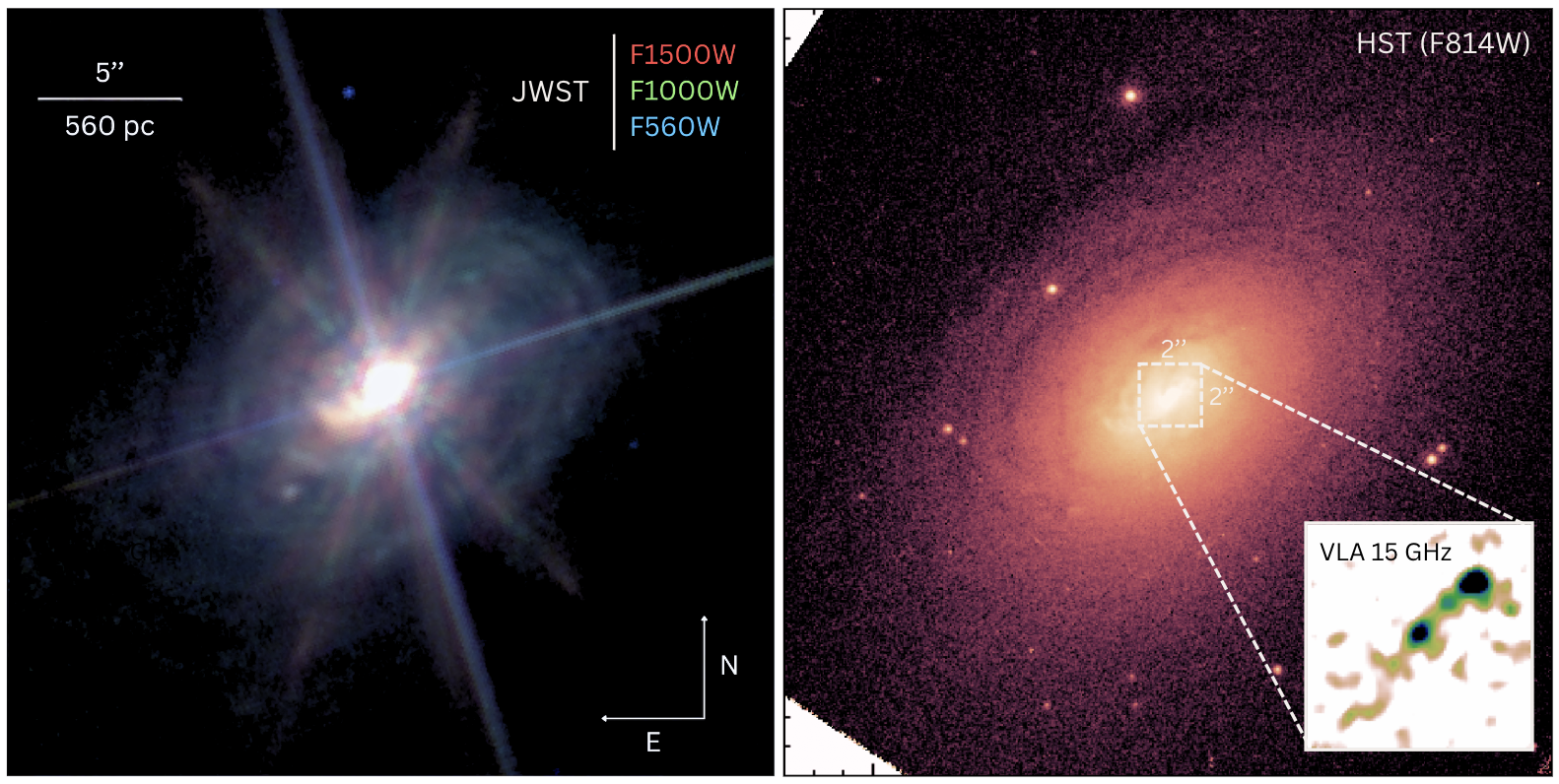}
    \caption{\textit{Left}: RGB composite image of the JWST/MIRI images of ESO 428-G14, where red, green, and blue channels correspond to $\rm F1500W$, $\rm F1000W$, and $\rm F560W$ filters, respectively. 
    The combination of these three filters reveals the MIR structure of the circumnuclear disk along with the small-scale nuclear extensions. The field of view (FOV) is $\sim 25''$ across. 
    The strong diffraction spikes are due to the bright central point-like source, which has not been subtracted in the version of the images used for this composite. 
    \textit{Right:} Optical HST image of ESO 428-G14 in the F814W filter over the same FOV as the JWST colour composite from the left panel. Zooming into the inner $2''\times2''$ region, the radio jet traced by the VLA 15 GHz emission is displayed in the bottom right of the figure.
    }
    \label{original_rgb}
\end{figure*}

ESO 428-G14 is a  spiral galaxy (see Fig.~\ref{original_rgb}) with a redshift-independent distance of $ \rm D \approx 23.2$ Mpc (a projected physical scale of $\approx 112.5$ pc/arcsecond) and a redshift of  
$z = 0.0057$ (values extracted from \hyperlink{https://ned.ipac.caltech.edu/byname?objname=ESO+428-G14&hconst=67.8&omegam=0.308&omegav=0.692&wmap=4&corr_z=1}{NED}). It is classified optically as a Seyfert 2 galaxy \citep{2006A&A...455..773V}, is Compton-thick in X-rays \citep{1998A&A...338..781M}, and has an intrinsic $\rm 2-10 \, keV$ luminosity $L_{\rm AGN} = 3.6 \times 10^{41}\, \rm erg s^{-1}$ \citep{Levenson+06},
and an estimated BH mass of $M_{\rm BH} \sim  (1–3) \times 10^{7} \, \rm M_{\odot}$ \citep{Fabbiano+19}. 

Observations  by \cite{Ulvestad+89} 
uncovered a bent radio jet elongated along the galaxy's major axis. Hubble Space Telescope (HST)  emission-line images revealed that the NLR is bi-polar and asymmetric, forming a rough double-helical shape to the North-West, and a larger elongated curved structure to the South-East, which appears to be co-spatial with the radio morphology \citep{Falcke+96,Falcke+98}. \\

Analysis  conducted by \citet{Riffel+06} using the Gemini Near-Infrared Spectrograph Integral Field Unit (GNIRS IFU) over the inner 300 pc found that the jet is launched with a slight inclination relative to the galactic plane. \citet{Riffel+06} also identified strong correlations between the flux and kinematics of various emission lines, notably the [\ion{Fe}{II}]1.257~$\mu$m line, and the radio emission, suggesting that the radio jet is responsible in part for the observed outflows. This is further supported by spectral analysis on the central $4\arcsec \times 4\arcsec$ region that favours the presence of shocks triggered by the interaction between the radio jet and the interstellar medium (ISM) in driving some of these outflows \citep{May+18,Fabbiano+18a}.

Recent ALMA CO\,(2-1) observations and modelling \citep{Feruglio+20} revealed the presence of a bar or warped disk, potentially contributing to the Compton-thickness observed in X-rays  \citep{Fabbiano+17,Fabbiano+18a,Fabbiano+18b}. Additionally,  \citet{Feruglio+20}
detected a bi-polar molecular outflow along similar positional angle (PA) to the radio jet, and at distances $\rm \sim 700\, pc$ from both sides of the nucleus.
\citet{Feruglio+20} suggest potential interactions between the AGN wind and material in the galactic disk could lift up  CO material from the galactic plane, leading to the observed CO outflow. \\ 

 This paper is structured as follows. We describe the JWST/MIRI imaging observations and data reduction techniques in Section~\ref{sec:2}. Ancillary data  used in this study are presented in Section~\ref{sec:archival-data}. Our findings are detailed in Section~\ref{sec:3}, followed by a discussion in Section~\ref{sec:4}. Finally, we summarise  our findings and conclusions in Section~\ref{sec:5}.

\section{JWST Observations \& Data reduction}
\label{sec:2}
\label{data} 

ESO 428-G14 is part of a sample of eight nearby Seyfert galaxies from the JWST Cycle 1 programme ``Dust in the Wind'' (ID: 2064, PI: David Rosario) within the ambit of the  \textit{Galactic Activity, Torus, and Outflow Survey} (\href{https://gatos.myportfolio.com}{GATOS}) collaboration \citep[see also][]{GarciaBurillo+21,Alonso-Herrero2021,GarciaBernete+23}.

On the $\rm 21^{st}$ of October, 2022, ESO 428-G14 was observed in five broad-band filters (F560W, F1000W, F1500W, F1800W, F2100W, corresponding to $\rm \lambda_{central}= 5.6,10,15,18,$ and 21 $\mu$m, respectively). The SUB256 subarray was used ($28\farcs2$ on a side), along with a fast readout mode and the shortest allowed number of groups ($N_{\rm groups} = 5$) to minimise the saturation of the bright nuclear emission expected in the galaxy. A relatively large number of integrations ($N_{\rm ints} = 50$) and four dithered exposures yielded a final exposure time of 358 seconds in each filter.

The raw data were downloaded from the Mikulski Archive for Space Telescopes (\href{https://mast.stsci.edu/portal/Mashup/Clients/Mast/Portal.html }{MAST}), and processed using the JWST pipeline python package version 10.2 with CRDS reference files \texttt{jwst\_1097.pmap}. 
The following sections outline the new modifications we made to the default MIRI imaging pipeline flow to improve the final imaging quality, particularly in the treatment of saturation and for corrections to the absolute astrometry.

\subsection{Saturation}
Even though our detector readout mode was designed to minimise saturation, some of the ramps for the nuclear pixels were still saturated due to the exceptional brightness of this AGN in the MIR.
Of the ESO48-G14 images, the F1500W exposures showed 12 inner pixels around the nucleus that were flagged as saturated using the standard pipeline. Within JWST's up-the-ramp fitting, saturation is evident when fewer than two viable groups are accessible for a ramp fit. By default, the MIRI imager pipeline excludes the first group to bypass potential transients from the readout step between integrations (see more details in Appendix A). We display in Fig. ~\ref{saturation} the X- and Y-profiles of the $\rm F1500W$ image of  ESO 428-G14,  spanning 6" across the nucleus. Here, we compare the default pipeline with our modified approach.  We show that by turning off the first frame correction  for the saturated pixels, we gained an additional viable group, facilitating a more accurate ramp fit and subsequent flux estimation.

\subsection{Astrometry}
In the final stages of the pipeline reduction, absolute astrometry is established through the \texttt{tweak\_registration} step. This process employs \texttt{DAOStarFinder}, a python algorithm which relies on finding background stars and/or galaxies for image alignment. Given that the galaxy fills all of the field of view (FOV) in our images, this method proves ineffective. We thus followed another approach, where we corrected the astrometry by aligning the target centroid of our images with the  AGN's nuclear coordinates   derived from ALMA, leveraging its high-resolution and precision as a reliable astrometric reference.
By setting the nuclear position to be the peak in the ALMA  1.3 mm continuum (which emerges from very close to the SMBH), we derive a nuclear position of  $\rm RA = 07^h\, 16^m\, 31.26s$ and $\rm DEC = -29\degree \, 19'\, 28.85''$ (ICRS). \\

\subsection{Point spread function}
\label{pointsourcesubtraction}

From the  three-colour JWST image in Fig.~\ref{original_rgb} (left panel), it is evident that the point spread function (PSF) exhibits several notable effects. This includes a distinct cruciform pattern, and  diffraction spikes that are due to the hexagonal shape of the mirror segments.
Consequently, in order to uncover the underlying central ($\sim 4'' \times4''$) MIR structure, image enhancement techniques such as point source subtraction and/or deconvolution become essential (see \citealt{2024AJ....167...96L} for a recent comparison of existing deconvolution algorithms). 

We used the \texttt{WEBBPSF} package \citep{2014SPIE.9143E..3XP} from \href{https://www.stsci.edu/home}{STScI}  to generate PSF models at the detector position of the AGN for each of the four dithered images, and  in each band. \texttt{WEBBPSF} allows the use of optimised PSFs including time-dependent wavefront distortion models and appropriate source  spectral energy distributions (SEDs), both which we employ for this work. The individual PSF frames were drizzled into a single PSF model image for each band using the same image reconstruction parameters as the corresponding science images. This replicates the detector sampling effects present in the observations which can influence the final PSFs. The point source accounts for 75\% of the F560W flux at the nucleus, 57-61\% of the flux in the F1000W, F1500W and F1800W filters, and 35\% of the flux in the F2100W filter. We show the PSF subtracted images in the top panel of Figure ~\ref{SED_plot}. A more focused study on PSF modelling will be presented in Rosario et al. 2024, in prep.

\subsection{Flux calibration}
\begin{figure*}
    \centering
    \includegraphics[scale=0.37]{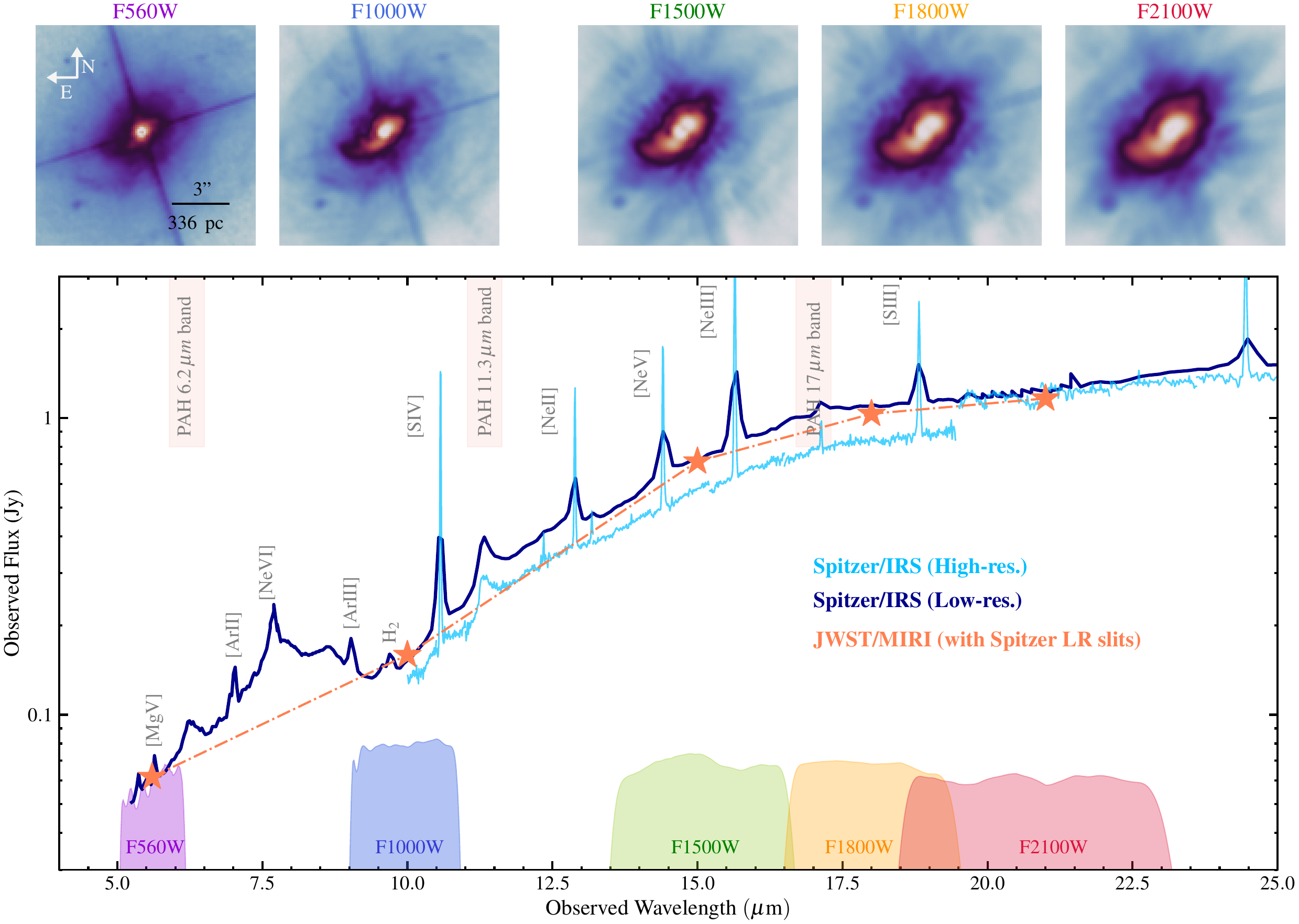}
    \caption{ \textit{Top:} PSF-subtracted JWST/MIRI images of ESO 428-G14 from the five filter bands. Each image spans a field of view of 11\arcsec$\times$11\arcsec, with North pointing up and East to the left. \textit{Bottom:} A comparison between JWST photometric measurements and existing {\it Spitzer} spectrophotometry. As a reference, key spectral features and PAH bands are identified. The depicted orange stars show JWST/MIRI aperture photometry, computed to mimic  Spitzer’s lower angular resolution at the respective wavelengths. This is achieved by convolving the images with Spitzer’s PSF and computing the expected flux within the spectral slits of the Spitzer/IRS low-resolution spectrograph. The fluxes obtained are in  good agreement with those from the Spitzer/IRS spectrum, an independent check on the accuracy of JWST/MIRI photometric calibration. The light blue curve shows the high-resolution Spitzer spectra, which show the main spectral features that lie within the JWST filter bandpasses (plotted along the bottom of the panel). Discrepancies in flux normalisation between the two Spitzer datasets are due to the different aperture extraction methods: optimal for high-resolution, and full for low-resolution.
    }
    \label{SED_plot}
\end{figure*}

The lower panel of Fig. ~\ref{SED_plot} presents a flux calibration check between JWST and calibrated spectroscopy from the Spitzer Space Telescope InfraRed Spectrograph (IRS; more details in Section ~\ref{spitzerirsdata}). For each filter, we convolved the JWST images with the Spitzer PSF\footnote{The Spitzer PSF information can be found in the IRS instrument handbook.} to simulate Spitzer's optical blurring effects. 
Subsequent aperture photometry was then performed to calculate the total JWST flux within the Spitzer-defined apertures, at each wavelength. 
This is shown in Fig. ~\ref{SED_plot} as orange stars, and is in good agreement with the SED of the Spitzer/IRS from both low-resolution (navy) and high-resolution (turquoise) spectra. The difference in normalisation between the two Spitzer/IRS spectra is a result of the different aperture extraction methods used. low-resolution data is presented using the full aperture extraction method, while high-resolution data is displayed using the optimal extraction method, which ensures a high signal-to-noise ratio and employs a smaller aperture size \citep[see][]{CASSIS}.

\section{Archival data}
\label{sec:archival-data}
We compiled data from several observations, which we describe in more detail below. The ALMA, HST, and VLA data were employed to  understand the structure and origin of the extended MIR emission within the broader context of this galaxy (see Section~\ref{multiwavelength-view}). Meanwhile, the VLT/SINFONI and the Spitzer/IRS data were used to compute and correct for emission line contamination (see Section~\ref{nature_MIRemission}).

\subsection{VLT/SINFONI spectroscopy}
K-band ($\rm 1.9-2.5\, \mu m$) IFU data were obtained with the Spectrograph for INtegral Field Observations in the Near Infrared (SINFONI) on the European Southern Observatory (ESO) Very Large Telescope (VLT) \citep{Eisenhauer2003,Bonnet2004}. The data were collected on December 22, 2010 and January 11, 2011, under Program 086.B-0484 (PI: Mueller Sanchez).  Observed in natural guide star mode, the data cover a roughly $\rm 4'' \times 4''$ field of view with a spectral resolution of $R \approx 1500$. A standard Object-Sky-Object observation sequence was used with each integration lasting 300 seconds for a total on-source time of 60 minutes.  Data reduction was carried out with the SINFONI Data Reduction Software. This data reduction pipeline included all standard routines needed for the near-infrared (NIR) spectra as well as additional routines used for construction of the data cube. Enhanced subtraction of the OH sky line emission was implemented as outlined by \citet{Davies2007}, and telluric correction and flux calibration was carried out using B-type stars.  The calibration uncertainty is about $\rm 10\%$ in flux as determined from the standard deviation of aperture photometries of the individual dithered cubes before combining to create a final data cube.

The emission line flux maps were generated using the custom IDL code LINEFIT \citep{davies2011}, which fits the emission line of interest with an unresolved line profile (a sky line) convolved with a Gaussian, as well as a linear function to the line-free continuum. This  single-component Gaussian  fitting procedure was performed for each spaxel of the data cube. Flux and kinematic information were extracted from the Gaussian fit and, in conjunction with the spatial information of each spaxel, used to generate the line maps for [\ion{Si}{VI}]$\lambda 1.96~\mu$m, Br$\gamma$ [2.16 $\mu$m] and H$_2$ 1-0 S(1) [$2.12\,\mu$m] emission lines. A more detailed analysis of the [\ion{Si}{VI}] and Br$\gamma$ lines is presented in \citet{May+18}.

\subsection{ALMA spectroscopy}
We extracted ALMA Band 6 data products for ESO 428-G14 from the \href{https://almascience.eso.org/aq/}{ALMA Science Archive} in rest frequencies of 230.538 GHz and 232.538 GHz corresponding to
the CO\,(2-1) line and the 1.33 mm continuum emission respectively. ESO 428-G14 was observed by ALMA on the $12^{th}$ of May, 2016 (progam ID: 2015.1.00086.S, PI: Neil Nagar). 
The FOV is $24\farcs52$  across, effectively covering the entire circumnuclear region. We processed the data using the \href{https://cartavis.org}{CARTA} software package and generated a flux map with a $2\sigma$ cut.  A more detailed  analysis of {these data was} conducted by \cite{Feruglio+20}.

\subsection{{\it HST} images}
We retrieved   fully reduced {\it HST}/WFPC2 images of ESO 428-G14  from  the \href{https://hla.stsci.edu/hlaview.html}{Hubble Legacy Archive} (progam ID: 5411, PI: Andrew Wilson). 
The observations were carried out on April 17, 1995, with exposures with the narrow-band F658N filter lasting a total of 13.33 minutes and exposures with the broad-band F814W filter lasting 3.33 minutes. We follow \citet{Falcke+98} and refer to the F658N image as the $\rm H\alpha$ image, despite the $33\%$ contribution from the [\ion{N}{II}]$\lambda\lambda6548,6584$ lines. A detailed analysis of this data was carried out by \citet{Falcke+96,Falcke+98}.

To correct the astrometry of the HST images, we  aligned it with the positions of stars identified from the GAIA DR3 catalogue \citep{collaboration2016description,GDR3}. These accurately identified stars within the HST images serve as precise astrometric reference points.

\begin{figure*}
  \centering 
   \raisebox{-\height + 5.6cm}{%
    \includegraphics[scale=0.355]{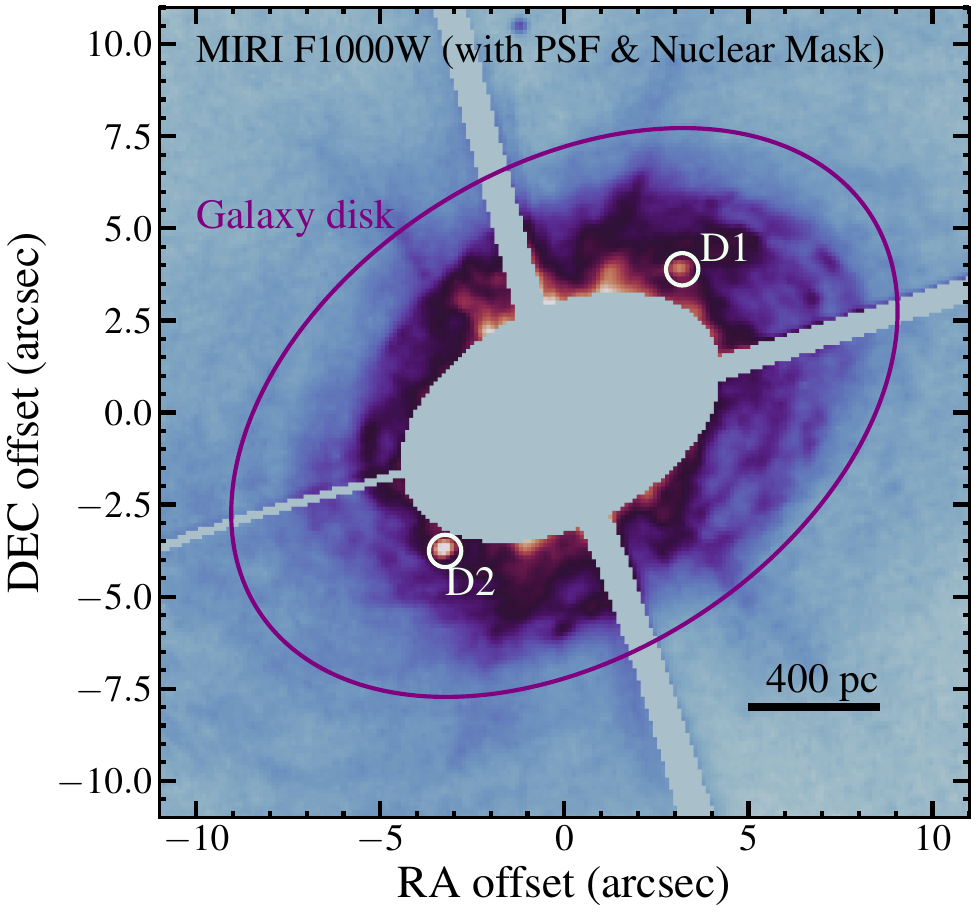} } 
    \includegraphics[scale=0.42]{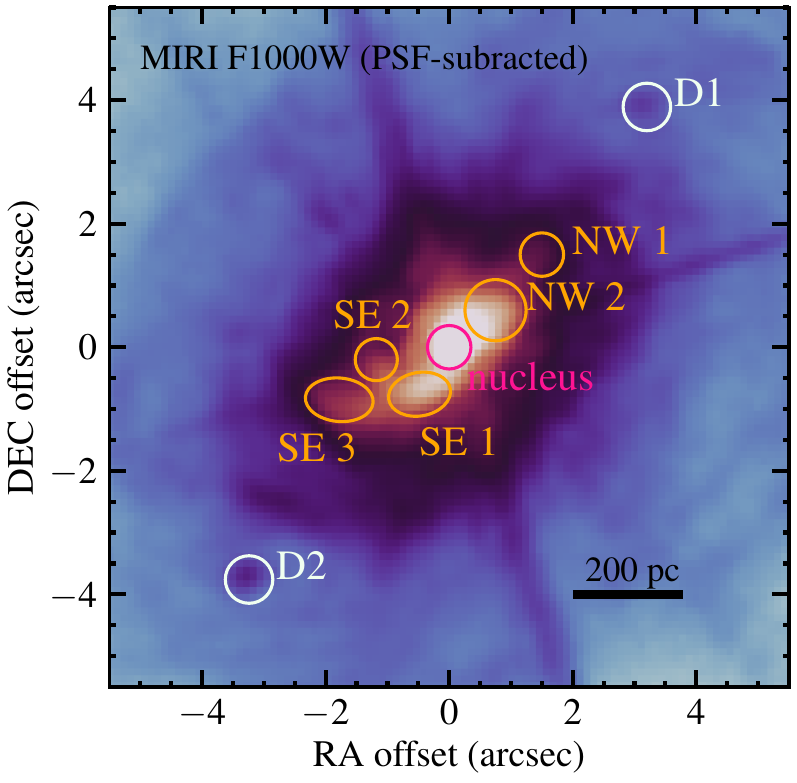} 
    \raisebox{-\height + 5.55cm}{\includegraphics[scale=0.335]{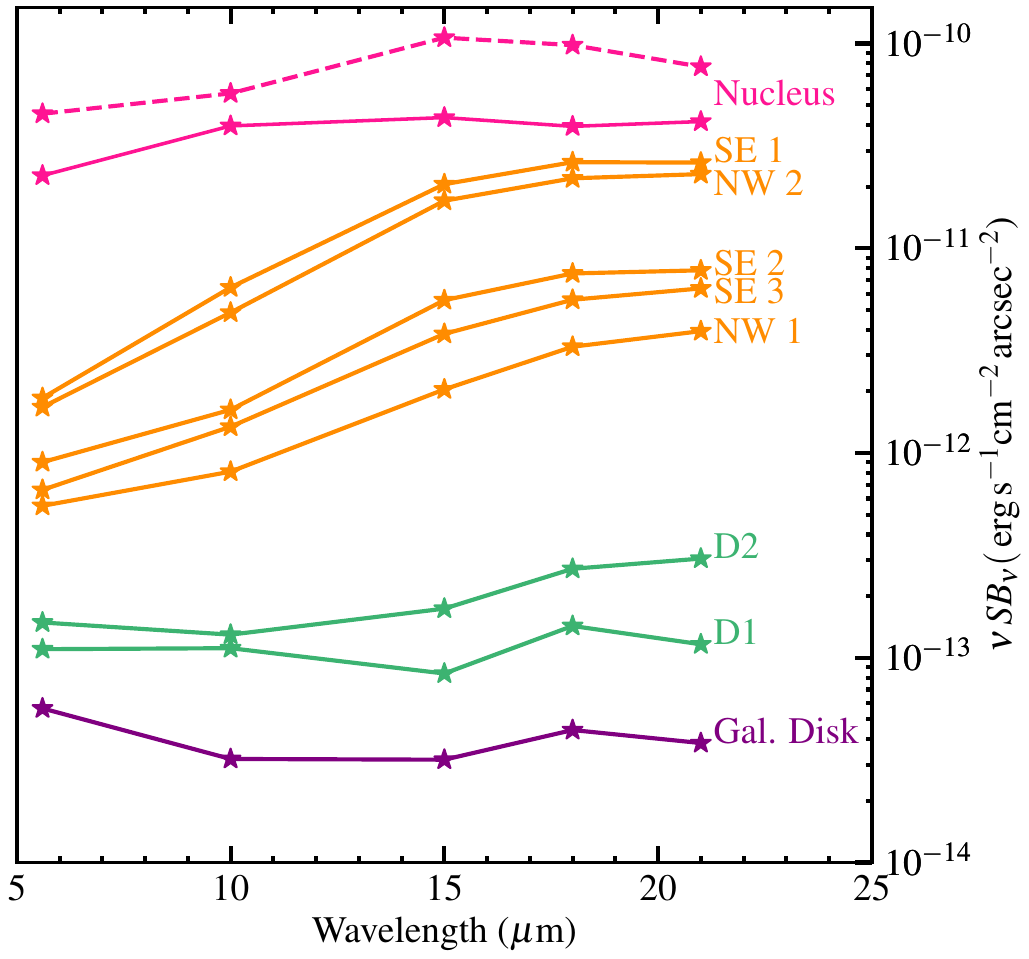} }
  \caption{ 
   \textit{Left:} The JWST/MIRI F1000W PSF-subtracted image showing the full galaxy disk and the aperture used to perform photometry on the disk emission (purple ellipse). The extended nuclear emission as well as the strongest diffraction spikes from the nucleus are masked out to enable a cleaner assessment of disk emission. Two bright regions on the disk are noted as D1 in the North West, and D2 in the South East (white circles).
   \textit{Middle:}
   A zoom into the inner $\sim 10''\times 10''$ region of the PSF-subtracted F1000W image to reveal the extended  MIR morphology. Regions of interest along the central 400 pc structure are marked in orange (NW1, NW2, SE1, SE2, SE3), while regions which appear within the disk are marked in  white (D1, D2).  
   \textit{Right:} Spectral energy distributions (SEDs) of the reduced surface brightness (SB), computed using PSF subtracted images across the JWST/MIRI bands for all regions, including the nucleus and the disk. For the nucleus, the dashed line is derived from the original images, while the solid line is produced using PSF-subtracted images. Note that the SB of the disk includes D1 and D2. Error bars are included in the plot but are too small to be visible. }
  \label{blob-SB}
\end{figure*}

\subsection{{\it Spitzer} IRS spectra} \label{spitzerirsdata}
We retrieved fully reduced low- and high-resolution MIR spectra (program ID: 30572, PI: Varoujan Gorjian) from the Combined Atlas of Sources with Spitzer/IRS Spectra \citep[\href{https://cassis.sirtf.com/atlas/query.shtml}{CASSIS};][]{CASSIS}. The spectra, taken in IRS staring mode,  were extracted 
using the optimal extraction method for high-resolution data to achieve the best signal-to-noise ratio, which is equivalent to a point source extraction. On the other hand, the low-resolution data were obtained using the full-aperture extraction method.

\subsection{VLA}
We extracted high-frequency (15 GHz, 2 cm) Very Large Array (VLA) U-band radio data from National Radio Astronomy Observatory data archive (PI: Heino Falcke). The radio core coordinates (RA, DEC) can be found in Table 3 of \citet{Falcke+98} and are $\rm RA = 07^h\, 16^m\,31.207^s$ and $\rm DEC = -29\degree \,19' \,28.89\arcsec$ (ICRS). The data  reduction 
was
performed as described in Section 2.2 of \citet{Falcke+98}. Astrometry corrections were done by placing the radio peak at the [\ion{Si}{VI}] flux peak.

\section{RESULTS}
\label{sec:3}

\subsection{Characteristics of the MIR morphology}
\label{sec:MIR_morphology}
  The new  JWST/MIRI images (Fig.~\ref{original_rgb})
provide a comprehensive view of the galaxy's various MIR characteristics, revealing new details in its centre. The image is dominated by a pronounced cross-shaped pattern of diffraction spikes, a manifestation of the very bright nuclear point-like MIR source spread out through JWST's intricate PSF. Both the JWST and HST images share a number of key features, including a circumnuclear dusty disk that appears as annular rings, which corresponds to faint, tightly-wound spiral arms.

These arms extend all the way to the nucleus, which appears prominently bright in both images in Fig.~\ref{original_rgb}. ESO 428-G14 harbours in its heart a well-collimated radio jet 
 (inset in right panel of Fig.~\ref{original_rgb} ) that is partially shaped by its interaction with the disk \citep{Falcke+96,Falcke+98}. \\

 The five single-band PSF-subtracted MIR mosaics revealing the resolved centre of ESO\,428-G14 are presented across the top of Fig.~\ref{SED_plot}. 
Residuals from diffraction spikes, particularly in the F560W filter, remain even after subtraction of the nuclear PSF, indicative of the high surface brightness central emission in the images. While the F560W image provides the best overall resolution, it does not reveal the nuclear extended structure prominently. This structure only emerges in the F1000W filter and in the longer-wavelength bands. However, in the images in these latter bands, the resolution gets steadily worse as the PSF broadens. 
We thus use the F1000W image as a baseline to study the nuclear extended structure, as it provides a good compromise between resolution and the visibility of the extended MIR features. For the remainder of the paper, we refer to the extended MIR structure seen in the central $4 \arcsec \times 4 \arcsec$, after subtraction of the nuclear point source, as the ``extended MIR emission''. \\

We find that the extended MIR emission exhibits a notable asymmetry, with distinct North-Western and South-Eastern regions extending along a similar PA to the radio jet of $\sim 131$ degrees, measured North to East. 
To the North-West, the emission adopts a compact structure and extends up to $\rm \sim  150 \, \rm pc$ from the nucleus. The South-Eastern emission is more extended, revealing two strands that diverge from the main structure, pointing northward. These can extend up to $\sim 300\, \rm pc$ from the nucleus. The total extent of the extended MIR emission is $\sim 450\, \rm pc$. As we will show in Section ~\ref{multiwavelength-view}, the overall structure resembles that of the H$\alpha$ line and the radio jet from \citet{Falcke+96,Falcke+98}.

\begin{table*}
\centering
\caption{The selected regions of interest as outlined in Fig.~\ref{blob-SB}.}
\label{tab:blob_table}
\setlength{\tabcolsep}{5.3pt}
\renewcommand{\arraystretch}{1.5}
\begin{tabular}{@{}l S[table-format=1.2] S[table-format=1.2] l S[table-format=3.2] S[table-format=3.2] S[table-format=3.2] S[table-format=3.2] S[table-format=3.2] S[table-format=2.2]@{}}
\toprule
Region$^{\rm \, (1)}$ & {$\rm \Delta\, RA$$^{\rm \, (2)}$} & {$\rm \Delta\, DEC$$^{\rm \, (3)}$} & Aperture$^{\rm \, (4)}$ & {$F_{\rm F560W}$} & {$F_{\rm F1000W}$} & {$F_{\rm F1500W}$} & {$F_{\rm F1800W}$} & {$F_{\rm F2100W}$} & {\%([\ion{S}{iv}]/F1000W)$^{\rm \, (5)}$} \\
\midrule
Nuclear$^{\rm \, (a)}$ & 0 & 0 & $0\farcs35$ & 85.31 & 189.40 & 536.15 & 588.54 & 534.53 &  \\
Nuclear$^{\rm \, (b)}$ & 0 & 0 & $0\farcs35$ & 16.00 & 51.00 & 84.00 & 91.00 & 112.00 & 6 \\
NW1 & 1.46 & 1.38 & $0\farcs35$ & 0.41 & 1.04 & 3.99 & 7.70 & 10.62 &  \\
NW2 & 0.97 & 0.47 & $0\farcs5$ & 2.42 & 12.39 & 65.92 & 101.31 & 122.91 & 30 \\
SE1 & -0.56 & -0.72 & $(0\farcs50, 0\farcs35, -80^\circ)$ & 1.95 & 11.95 & 57.72 & 88.27 & 101.72 & 15 \\
SE2 & -1.51 & -0.38 & $0\farcs35$ & 0.65 & 2.10 & 10.89 & 17.56 & 21.06 & 35 \\
SE3 & -2.08 & -1.00 & $(0\farcs55,0\farcs35,-95^\circ)$ & 0.75 & 2.71 & 11.65 & 20.42 & 26.85 & 40 \\
D1 & 3.44 & 3.89 & $0\farcs4$ & 0.096 & 0.17 & 0.19 & 0.39 & 0.37 & \\
D2 & -3.85 & -3.76 & $0\farcs4$ & 0.13 & 0.20 & 0.40 & 0.75 & 0.98 & \\
\bottomrule
\end{tabular}
\\
\small  \textit{Notes --}  (1)  Regions of interest (refer to Fig.~\ref{blob-SB}).  For the nucleus, we present two SEDs, one derived from (a) the original images (dashed pink line in Fig.~\ref{blob-SB}), and the other from (b) the PSF-subtracted images (solid pink line, Fig.~\ref{blob-SB}). For all the other regions, the SEDs are derived using PSF subtracted images. (2\&3)  The relative positions are given as offsets from the nucleus  ($\rm RA_{\rm nuc} = 07^h\, 16^m\, 31.26^s$ and $\rm DEC_{\rm nuc} = -29\degree \, 19'\, 28.85''$) in arcseconds. (4) The aperture radius is provided in arcseconds, showing a single value for circular apertures and three parameters (semi-major axis, semi-minor axis, PA measured N-E) for elliptical apertures. (5) When possible, we provide the average percentage of line emission contribution to the total flux (in the case of the F1000W) filter. All fluxes across various filters are measured in  units of $\rm 10^{-3}\, Jy$.
\label{table_blobs}
\end{table*}

\subsection{MIR regions of interest} \label{regions}
We leverage the optimal resolution and sensitivity of the F1000W image
to define regions of interest within the nuclear and extended MIR structure, as depicted in the middle panel of Fig.~\ref{blob-SB}. We then examine and compare the MIR spectral energy distributions of these regions, measured in fixed sized apertures, to explore their properties. The regions and details of the apertures are given in Table \ref{tab:blob_table}.

The nuclear aperture is centred on the nuclear position.
NW1 \& NW2 are situated in the North-West region, while SE1, SE2 \& SE3 map out the two strands that split from the main structure towards the South-East. 

\begin{figure*}
  \centering
  {\includegraphics[width=1.07\textwidth]{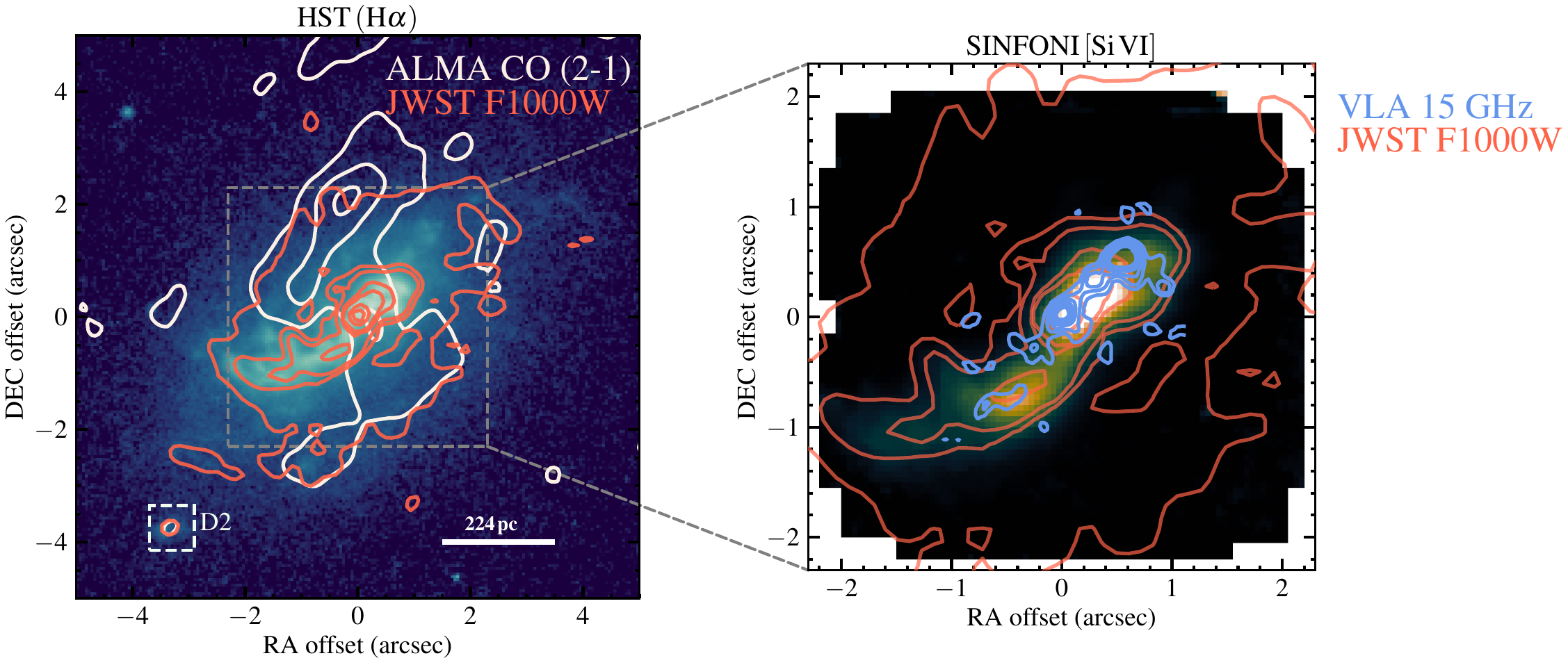}}
  \caption{ 
  \textit{Left:} The HST $ \rm H\alpha$ image of ESO 428-G14, with the nucleus at the centre. Superimposed red contours depict the JWST PSF-subtracted F1000W image, while the white contours are for the ALMA CO\,(2-1) moment 0 map. The position of the SE region D2 (see Fig.~\ref{blob-SB} as a reference) is highlighted by the white square. \textit{Right:} Zooming into the inner $\sim 4''\times 4''$ to match the FOV of SINFONI, we display the [\ion{Si}{VI}] flux map. The red contours are from the JWST PSF-subtracted F1000W image (comparable to the left panel), while light blue contours map the VLA radio 2 cm (15 GHz) image. 
  }
  \label{jwst_plus_anci}
\end{figure*}

We also identified 
two interesting regions, D1 ($07^h 16^m 30.97^s$, $-29^\circ 19' 24.90''$) and D2 ($07^h 16^m 31.46^s$, $-29^\circ 19' 32.55''$), that lie beyond the extended MIR emission, but along a similar PA to it and to the radio jet. Unlike most of the galaxy disk, these regions appear prominently red in the map of the colour F560W-F1000W (see Fig.~\ref{f560W-f1000W-ratio} in the Appendix). Their location and relationship to the extended MIR emission suggests that the AGN could play some part in their heating. Indeed, recent work by \citet{HermosaMunoz2024} 
detect multiple MIR knots along the direction of the outflow in another nearby Seyfert galaxy, NGC 7172. 
The left  panel of  Fig.~\ref{blob-SB} zooms out to show the full circumnuclear disk. We define an aperture mask that excludes any contribution from the nuclear or extended MIR emission, as well contamination from  diffraction spikes from the nuclear point source. This aperture gives an averaged SED for the circumnuclear disk free of much emission influenced strongly by the AGN.\\

For all identified regions, we select apertures with diameters bigger than the PSF FWHM of the longest wavelength filter (i.e. F2100W, FWHM = $0\farcs67$). This minimises the need to include detailed aperture corrections in the SEDs with increasing wavelength (see Table \ref{table_blobs}). The right panel of Fig.~\ref{blob-SB} shows the five band SEDs, expressed in units of reduced surface brightness ($\nu\, \rm SB_{\nu}$), of these different regions. Across all filters, regions closer to the nucleus and within the direct influence of the AGN, are characterised by high SB. The regions farther out, identified with the galaxy disk, exhibit lower SB. \\

The highest SBs are found in the nuclear aperture (pink line), particularly at shorter wavelengths ($\rm \lambda_{cent} \leq 10\, \mu m$). For completeness, we also add in Fig.~\ref{blob-SB} and Table \ref{table_blobs} the nuclear SED derived from the original images (dashed pink line), which includes the PSF. This shows a peak around $\sim 15\, \rm \mu m$. The nuclear SED derived using the PSF subtracted images (solid pink line) is flat,  with a $\rm SB_{F560W}/SB_{F2100W}$  ratio of $\sim 0.56$, and remains consistently bright
across all filters.  This indicates that the nuclear region emits a substantial portion of energy ($\sim 20\% $ of the total F1000W flux), which does not significantly change at longer wavelengths.  This flat SED can be attributed to dust with a blend of temperatures from various components in the vicinity of the AGN, out to $\sim 40$ pc, reflecting a complex thermal environment rather one with a single uniform temperature. \\

Moving towards regions in the extended emission structure, we find that
all regions identified within it (in orange) are characterised by steeper SEDs, defined by their $\rm SB_{F560W}/SB_{F2100W}$ ratios of $\sim [0.07-0.1]$. 
Lower values of this ratio indicate substantial increase in flux at longer wavelength, while values closer to unity would imply a flat SED. Assuming a simple black body, we estimate temperatures within the range $\sim 100-120\, \rm K$ for the regions in the extended emission. The highest temperatures are found in SE1 and NW2, likely due to their close proximity to the nucleus.  These values serve as an approximation (based on five photometric points), and a more advanced fitting method is required to accurately estimate the temperature. \\

In constrast to the extended emission regions influenced by the AGN, the average galaxy disk has a relatively flat SED. The rise we observe towards the shortest wavelengths could come from low levels of stellar light in the MIR, the tail of the combined photospheric emission from stars in the disk. A slight excess in the F1800W band over a flat SED shape could arise from the contribution of Polycyclic Aromatic Hydrocarbon (PAH) bands at 17 microns (indicated in Fig.~\ref{SED_plot}; see \citet{2022A&A...666L...5G} for a detailed discussion). 
We interpret the flatness of the star-forming galaxy disk in a similar way as we do for the AGN nucleus: the disk consists of amalgamated dusty star-forming regions with a range of temperatures set by distributed heating. Because stars are not as powerful heating sources as the AGN, the SB of the disk is low.

Turning to the 
regions D1 and D2 that appear to be on the galaxy disk (although might also lie along the outflow direction),
these display flatter SB profiles similar to the galaxy disk. Their SB ratios of  $\rm SB_{F560W}/SB_{F2100W}$ $\approx [0.7-0.95]$ highlight the flatness of these SEDs. For this reason, we suggest that these features are star-forming regions. While both D1 and D2 share this characteristic, both show minor differences in SEDs, indicating variation in the star-formation heating.
The elevated surface brightness in the F1800W band is especially apparent in region D1, which again suggests a possible PAH contribution. Estimating a temperature for these SF regions would require accounting for PAH, which is beyond the scope of this paper.  \\

\subsection{A multi-wavelength view}

\label{multiwavelength-view}

\begin{figure*}
    \centering
    \includegraphics[scale=0.4]{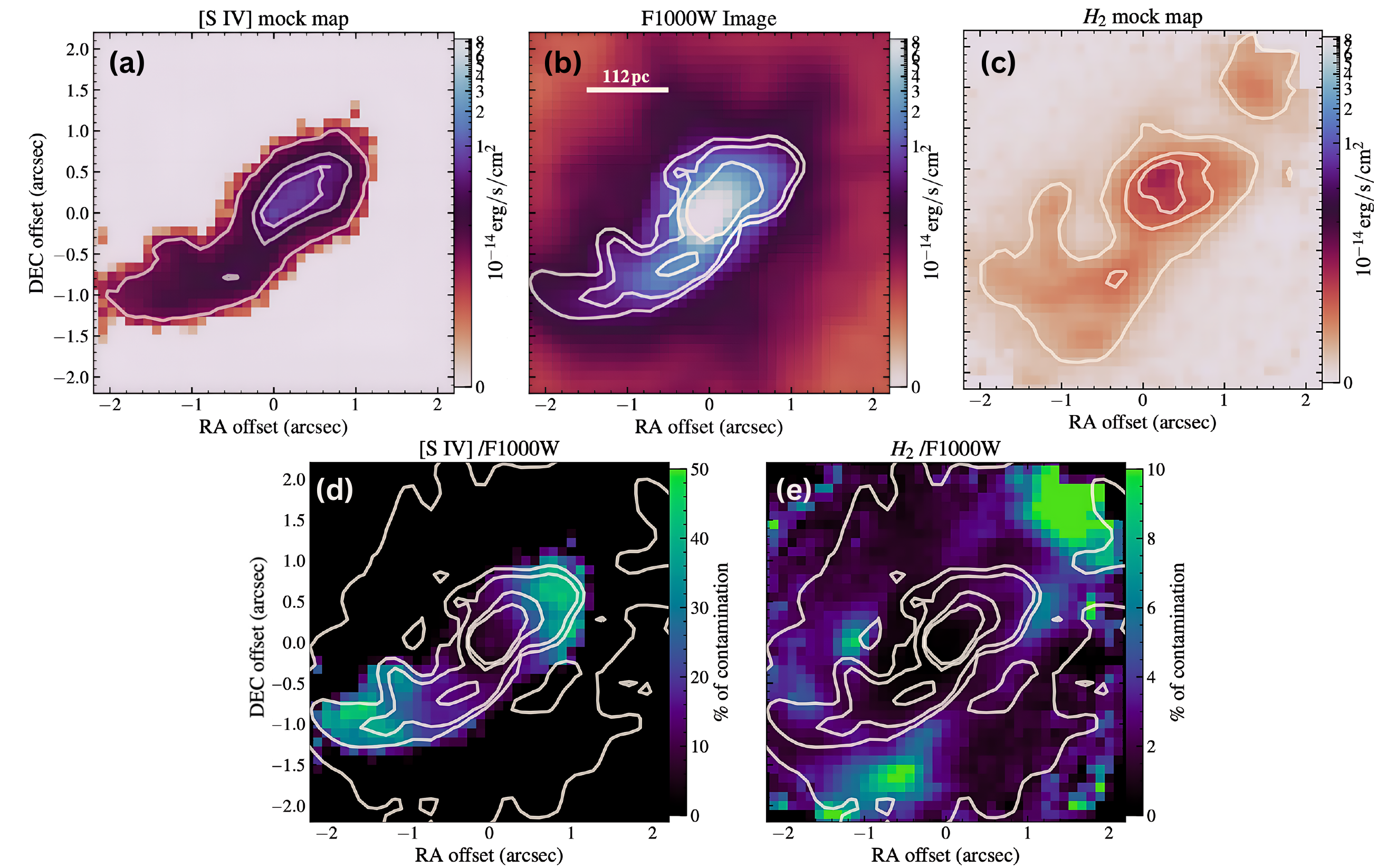}
    \caption{Emission line mock maps and the associated contamination maps with respect to the F1000W image. The top three panels share the same logarithmic scale and stretch, allowing for a direct visual comparison of the relative flux. For each panel, contour levels represent (2,10,50) $\sigma$ levels.  (a) Mock [\ion{S}{iv}]$\lambda 10.51\,\rm \mu m$  map computed using Br$\gamma$ from SINFONI as a template. (b) The PSF-subtracted F1000W JWST/MIRI image. (c) Mock $H_{2}$ 0-0 S(2) [$9.65\, \rm \mu m$] map using $H_{2}$ 1-0 S(1) [$2.12\, \rm \mu m$] from SINFONI as a template. (d) Spatial distribution of contamination percentages of [\ion{S}{iv}] within the F1000W image.  Overlaid contours are from  the JWST image to show  the extent of the MIR morphology. (e) Same as before but for $H_{2}$ contamination level.}
    \label{contamination_map}
\end{figure*}

The most remarkable find from  Fig.~\ref{jwst_plus_anci} is the 
striking morphological resemblance between the MIR emission, the HST  $\rm H\alpha$ image and the [\ion{Si}{VI}] emission.  This agreement may suggest that the processes influencing the ionised gas  and the MIR emission (arising from dust) in the nuclear region may be the same, or that the MIR emission is not solely attributable to dust (see next Section \ref{nature_MIRemission}). \\

In the left panel of  Fig.~\ref{jwst_plus_anci}, the CO\,(2-1) structure appears to trace an inner feature near the vicinity of the SMBH. \citet{Feruglio+20} describe this
as
a transverse gas lane that is bridging two segments of a circumnuclear ring, and feeding the SMBH by channeling material to it.
Our observations reveal  both the MIR structure and $\rm H\alpha$ emission to be perpendicular to 
this ``feeding'' lane, implying a polar configuration. This is also supported by the alignment of the radio jet  with the extended MIR structure, as discussed in Section~\ref{sec:MIR_morphology}.
This multi-wavelength view uncovers that D2 is also detected in H$\alpha$ and CO\,(2-1) emissions (see left panel, Fig.~\ref{jwst_plus_anci}), strengthening the case for its SF nature. \\

The radio morphology, with an extent of $\sim \rm 2''$ along PA$\approx 131 \degree$,
reveals more intricate details, especially towards its hot spot in the North-West. Here, it captures  two clumps that are not resolved in the JWST/MIRI images (see NW2 in Fig.~\ref{blob-SB}). 
This feature has been previously identified in  HST $\rm H\alpha$ imaging by \cite{Falcke+96} and in the SINFONI Br$\gamma$ map  by \cite{May+18}.

\subsection{Emission line contamination}
\label{nature_MIRemission}

JWST/MIRI images are obtained using broad-band filters that span a few microns around the central wavelength.
This means that the filters can capture both the continuum emission from dust and any line emission from gas that falls within the filter's wavelength range. As can be seen in the bottom of Fig.~\ref{SED_plot},  the five  imaging filters used in this study overlap with a couple of emission lines and PAH bands that could contribute to the MIR emission, potentially influencing the morphology seen.  This includes the F1000W filter, which can be contaminated by  the [\ion{S}{IV}] and $\rm H_{2}$  0-0 S(3)  lines; F1500W,  by  [\ion{Ne}{V}] and [\ion{Ne}{III}] lines; and F1800W, where PAH bands and the [\ion{S}{III}] line may impact the observed emission.
Moreover, the mutli-wavelength analysis in Fig. ~\ref{jwst_plus_anci} shows a striking correlation between the MIR structure, [\ion{Si}{VI}], and H$\alpha$ emissions, which could be explained (in part) by strong emission line contribution in the JWST images. 
This could influence our interpretation of the observed MIR structures,  attributing features to dust continuum that may, in fact, be closely connected to ionised or molecular gas instead.  \\

The absence of any high-resolution JWST spectroscopy for ESO 428-G14, such as from the MIRI Mid-Infrared Spectrograph (MIRI/MRS), complicates this analysis, as it would otherwise provide a clearer distinction between dust and gas contributions to the observed MIR structure. To overcome this limitation, we leverage the wealth of multi-wavelength information for this galaxy to develop a method to estimate the level of emission line contamination in the F1000W image. As discussed before, this filter provides the highest resolution of the inner ($4 \arcsec \times 4 \arcsec$) extended MIR structure, and is thus the ideal band for our assessment of contamination along that region. \\

\begin{figure*}
    \centering
    \includegraphics[scale=0.5]{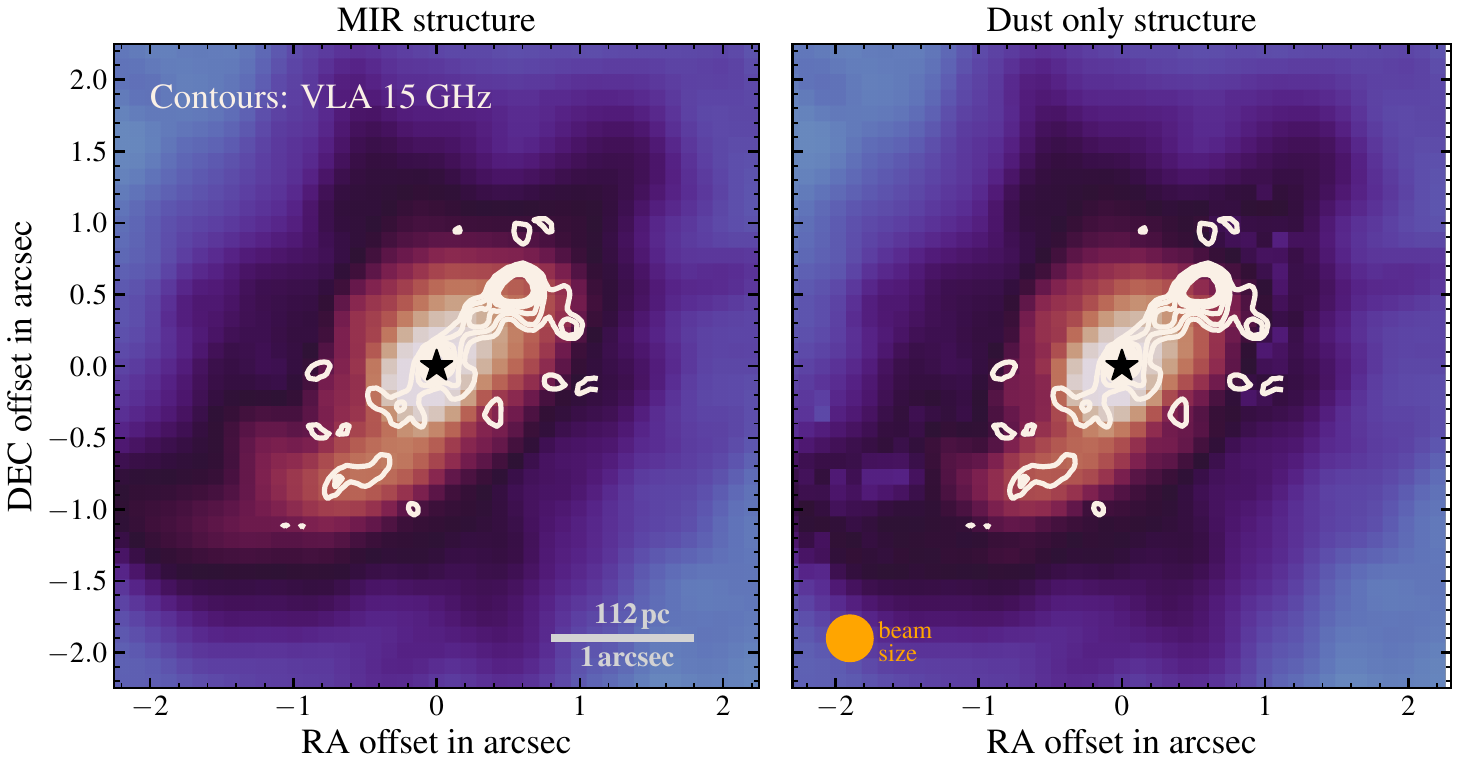}
    \caption{\textit{Left:} The original polar and extended MIR structure from the PSF-subtracted F1000W image.  Contours show the VLA 15 GHz emission which traces the radio jet.  The black star marks the position of the nucleus. \textit{Right:} The pure continuum or ``dust-only'' image obtained after subtracting the emission line contributions from the PSF-subtracted F1000W image. The dusty structure shows good spatial correspondence to the radio jet, including an asymmetry in the NW extent compared to the SE extent.  The orange circle in the bottom left shows the FWHM ($0\farcs32$) of the F1000W PSF.  Both images are displayed on a logarithmic scale and share the same stretch to highlight the differences before and after decontamination.} 
    \label{dustonly-image}
\end{figure*}

From Fig.~\ref{SED_plot}, one can see that both the [\ion{S}{IV}] line at $\rm \, 10.51\, \mu m$ and the $\rm H_{2}$ line at $\rm \, 9.65\, \mu m $ fall within the F1000W filter bandwidth, which means they can contaminate this image.
To check this, we produced mock maps for $\rm H_{2}$ and [\ion{S}{IV}]. Our methodology can be applied to any filter and is described as follows. First, 
we leverage  the Spitzer/IRS SH spectrum (when possible, otherwise low spectral resolution) to estimate the total flux due to each line. Second, we use these estimates to scale a SINFONI line map of comparable ionisation potential (IP) to ensure similarity in morphological characteristics.
This then produces a mock map for the line in question, providing us with a proxy for its expected spatial distribution. \\

It is important to note that the ionisation potential can significantly influence the morphology of the emission line.  A recent JWST study by \citet{Pereira-Santaella+22} demonstrate that MIR lines with higher IPs tend to be more compact compared to those with lower IPs.  Another caveat that should be kept in mind is extinction, which is very low for this object and thus has no impact on the measurements made.
Nevertheless, this method allows us to estimate the extent of emission line contribution within the JWST F1000W image, thereby revealing regions with significant contamination. Our  method has been validated against another nearby Seyfert from the GATOS survey, NGC~5728, for which we have both imaging and spectroscopy (see upcoming sample paper Rosario et al. 2024, in prep). We describe in more detail below the steps taken to produce each mock map.

\subsubsection{[\ion{S}{IV}] mock map}
Using  Spitzer/IRS SH spectrum (see  Fig.~\ref{SED_plot}), we fit the [\ion{S}{IV}] emission line   with a single-component Gaussian, yielding 
a total flux of $f_{\rm SIV}= 7.564 \times 10^{-13}\, \rm  erg/s/cm{^2}$.
To generate a mock [\ion{S}{IV}] map, we used the Br$\gamma$ map as morphology template.  \citet{Riffel+06} demonstrated that Br$\gamma$ is in good correspondence with the radio jet, with some signatures of shocks. As such, we assume that there is little to no SF contribution in this line. We also note that Br$\gamma$  has an IP of 13.6\,~eV that can be  comparable to that of  [\ion{S}{IV}] (35\,~eV), so the extent of their emissions will not differ by much.  The mock map is then produced by scaling the Br$\gamma$ map  to match the expected flux from [\ion{S}{iv}], $f_{\rm SIV}$. \\

The level of [\ion{S}{IV}] contamination is quantitatively assessed by calculating the ratio between the mock [\ion{S}{IV}] map (panel a in Fig.~\ref{contamination_map}) and the PSF-subtracted JWST F1000W image (panel b in Fig.~\ref{contamination_map}). 
The ratio map, as seen in Fig.~\ref{contamination_map} (panel d),
unveils how the [\ion{S}{IV}] contamination level varies spatially across the extended MIR morphology. Overall, the median contribution of [\ion{S}{IV}] across the  entire  extended MIR structure is approximately 15\%.
Notably, structures extending $\sim 1''$ from nucleus, both towards the NW and SE regions, are the most impacted, with contamination levels reaching up to 45\%.\\

\subsubsection{$\rm H_{2}$ mock map}
We extracted from \citet{Spoon+22} the total flux arising from the $\rm H_{2}$  0-0 S(3) [$\rm \, 9.65\, \mu m $] line as $F_{\rm tot, \rm H_{2}}= (9.2\pm1.6) \times 10^{-21} \, \rm W~cm^{-2}$, computed using Spitzer/IRS low-resolution spectral data.
With this value, we scaled the SINFONI $H_{2}$ 1-0 S(1) [2.12 $\mu$m] map 
to produce a
mock map for $\rm H_{2}$  0-0 S(3) [$\rm \, 9.65\, \mu m $] line (panel c in Fig.~\ref{contamination_map}).  
We then computed a contamination map as a ratio between the mock $\rm H_{2}$  0-0 S(3) [$\rm \, 9.65\, \mu m $] map and the F1000W image (panel e in Fig.~\ref{contamination_map}). This gives a very weak contribution of  $\rm H_{2}$  0-0 S(3) [$\rm \, 9.65\, \mu m $] line to the extended MIR emission, varying between 2\% to 4\%.  \\

\subsubsection{Pure dust continuum morphology}

To reveal the true morphology of the dust, we subtracted the total flux contribution arising from both the [\ion{S}{IV}] and $\rm H_{2}$ from the F1000W PSF subtracted image.
From the maps computed in Fig.~\ref{contamination_map}, it is clear that  $\rm H_{2}$ contributes negligibly to the total MIR emission, with a contamination level of 4\% at its highest. In contrast, the [\ion{S}{IV}] line accounts for nearly half of the extended emission, establishing it as the primary contaminator. Consequently, we boosted the [\ion{S}{IV}] flux by $4\%$ to account for the $\rm H_{2}$ contribution.  Following this, 
we subtracted the boosted [\ion{S}{IV}] map  from the PSF-subtracted F1000W image and produced a contamination free image, as shown in  right panel of Fig.~\ref{dustonly-image}. 
This image brings to light the structure of the ``pure continuum''.  Indeed, the brightness of the MIR structure significantly drops in the SE and NW. This affects the appearance of the South-Eastern strands which are now too faint to notice. As a result, the structure of the dust appears more compact and concentrated around the nucleus. When compared to the regions of interest presented in Fig.~\ref{blob-SB} and Table~\ref{tab:blob_table}, it becomes evident that contamination levels rise with increasing distance from the nucleus. For example, the South-Eastern regions, SE2 and SE3, exhibit contamination levels ($\sim$35\% and 40\%) at least twice as high as that of SE1 ($\sim 15\%$), resulting in their diminished appearance in Fig. ~\ref{dustonly-image}. \\

Compared to the original polar and extended MIR structure presented in left panel of Fig.~\ref{dustonly-image}, the new ``dust only'' image  shows that it reduced by more than half of its original size. The new total extent of the dust is roughly  $\sim 1.8''$  ($\sim 200$ pc) in size, stretching  $\sim 0.8''$ to the NW and $\sim 1''$ to the SE from the nucleus.  However, it still exceeds the traditional limits of the torus ($\sim$ tens of parsecs). Moreover, we find a striking correspondence between the extent and morphology of  the polar dust and the radio jet (see right panel, Fig.~\ref{dustonly-image}). We explore in Section~\ref{sec:4} all possible origins of this polar dust. \\

\section{DISCUSSION}
\label{sec:4}

\subsection{Dust beyond the torus}
The exceptional resolution and sensitivity of JWST/MIRI allowed, for the very first time, the imaging of the MIR heart of ESO 428-G14, revealing new details about its structure and properties.
In the central $r \approx 2\arcsec$ region of ESO 428-G14, we detect a bi-polar structure extending  up to $\sim 450$ pc in size, from the NW 
to the SE.
In contrast to the galaxy disk,  our SED analysis reveals that regions along the polar MIR structure emit strongly in the MIR,  and are characterised by warm temperatures ($T_{\rm dust} \approx 120\, \rm K$) and steep SEDs ($\rm SB_{F560W}/SB_{F2100W} \approx 0.07$), especially for regions (NW2, SE1) that are adjacent to the nucleus. Such trends in the MIR strongly suggest the presence of warm dust along these regions.  \\

Our multi-wavelength analysis (Fig.~\ref{jwst_plus_anci}) points towards a strong correlation between the structure of the extended polar MIR emission and that of multiple emission lines from ionised gas ([\ion{Si}{VI}], H$\alpha$, Br$\gamma$).
We attribute this, in part, to a significant contribution of emission lines in JWST/MIRI broad-band imaging that contaminate the MIR morphology (Section ~\ref{nature_MIRemission} and Fig.~\ref{contamination_map}). While it is challenging to disentangle  gas and pure dust structures in the absence of MIR spectroscopy (e.g. from the JWST/MIRI MRS), an important aspect of this study is the development of a method to estimate the level of line contamination using existing pre-JWST datasets (see Section~\ref{nature_MIRemission} for more details). 
Using the image in the F1000W filter, an optimal band for the contrast of extended MIR emission, we produced a pure continuum version that reveals that the dust is indeed more compact, with a total length of  $\sim 200$ pc  (Fig.~\ref{dustonly-image}) - shrinking to half of its original  length ($\sim 450$ pc). \\

The question now arises regarding the nature of this polar dust. 
\citet{Prieto+14} computed  an extinction map for ESO 428-G14 (see their Fig. 8), revealing dust structures with weak spirals which are reminiscent of the morphology seen in CO\,(2-1) emission from ALMA \citep[][their Fig. 1]{Feruglio+20}. These structures are indeed different from the polar structure seen in Fig.~\ref{dustonly-image}. As such, the dust probed with JWST  is not normal cold dust in the galaxy, but it could be dust in the NLR illuminated by AGN radiation fields, local dust heated by jet-induced shocks, or dust driven by a wind from the torus. We discuss each of these possible scenarios in more detail.

\subsection{Dust illuminated in the NLR}
\label{dust_illumination}

The simplest explanation of the polar dust would be the illumination of ambient dust within the NLR by the AGN's radiation. Such a scenario accounts for the observed spatial correlation with ionised emission line regions because both phases are produced by the same radiation field.
This explanation does not require any connection between the dust and outflows from the AGN. Here, we test whether heating/illumination from the central engine alone is able to explain the observed temperatures along the polar MIR emission. 

As an exercise, we take the region NW2 (Fig.~\ref{blob-SB}) as a case study, at a distance $D \approx 100 \, \rm pc$ from the nucleus.  To compute a bolometric luminosity for the AGN, we applied a bolometric correction according to Eq. 3 in \citet{2020A&A...636A..73D} to the hard X-ray luminosity of $L_{2-10\, \rm keV}= 3.6\times 10^{41}\, \rm erg\, s^{-1}$ reported by \citet{Levenson+06}. This gives $L_{\rm bol, \, AGN} = 5.5\times 10^{42}\, \rm erg\, s^{-1}$. This value is comparable to the estimate of the AGN's bolometric luminosity from its integrated MIR luminosity  of $ 2 \pm 1 \times 10^{42}\, \rm erg\, s^{-1}$ from \citep{May+18}.

We express the expected anisotropic radiation field in terms of the Habing field, following \citet{Tielens+10} (Chapter 5, Eq. 5.43): 
\begin{equation}
    G_0 = 2.1 \times 10^4 \left( \frac{L_{\rm bol, AGN}}{10^4 L_{\odot}} \right) \left( \frac{0.1\, \text{pc}}{D} \right)^2.
\label{eq:G}
\end{equation}

Then, for dust grains with a size $a$, the equilibrium dust temperature can be expressed as:
\begin{equation}
    T_d \approx 33.5 \left( \frac{1\, \mu m}{a} \right)^{0.2} \left( \frac{G_0}{10^4} \right)^{0.2} \, \rm K,
\label{eq:tdust}
\end{equation}
\noindent following \citet{Tielens+10} (Chapter 5, Eq.~5.44). An important assumption here is that the intrinsic AGN SED that heats the dust peaks strongly in the UV, roughly like the local Habing field.

Assuming classical ISM dust grains  within the range of $a= 0.005-0.25\, \rm \mu m$ \citep[][see their discussion around Eq. 5.1]{DraineLee1984}, the expected dust temperature at distance $D \approx 100\, \rm pc$ from central source with the assumed $L_{\rm bol, \, AGN}$  would be $T_{\rm dust} =[76, 35]\, \rm K $, where the lower temperature correponds to the largest grain size. Therefore, assuming the absorption-averaged grain size is at the lower end of size distribution, reasonable for ISM dust, illumination along the NLR can heat the dust only up to several tens of K. This is lower than the dust temperatures we estimate for the NW2 region. 

Note that the temperature may be a bit higher than our estimates if X-ray heating of dust is efficient, because AGN have a higher proportion of X-ray energy compared to the Habing field. Additionally, in the ionised gas phase, resonance trapping of Ly$\alpha$, which leads to complete reprocessing of these UV photons, can also raise the temperature of the dust associated with this phase.

\subsection{Dust originating from the torus}

At the inner edge of the torus, infrared radiation pressure applied on dust grains can lift the dust radially outward, leading to a polar dusty wind \citep[e.g.][]{Honig+12}.  
Semi-analytical models produced by \citet{Venanzi+20} confirm that it is possible for dusty winds to be launched from the inner regions of the torus. Their work also  predicts radiation pressure to be more efficient at driving this wind around a critical limit, where AGN radiation pressure and gravity from the black hole are balanced. \\

Some studies have proposed that the column density - Eddington ratio ($\rm N_{H}-\lambda_{Edd}$) plane can be used as a diagnostic to infer the conditions under which radiation pressure dominates over gravity, thereby giving rise to  radiation-driven outflows \citep[e.g.][]{Fabian+08,Ricci+17,Venanzi+20,Alonso-Herrero2021}. For ESO 428-G14, \citet{Feruglio+20} estimated the line-of-sight $H_{2}$ column density to be $\rm N_{H_{2}} \approx 2 \times 10^{23}\, cm^{-2}$,  but report this as a lower limit due to beam dilution. This equates to a hydrogen column density of $\rm N_{H} \approx 4 \times 10^{23}\, cm^{-2}$. 
Using the $\rm M_{BH}-\sigma_{\star}$ scaling relation, \citet{Fabbiano+19} derived an estimate for the BH mass of $\rm M_{BH} \approx (1-3) \times 10^{7}\, M_{\odot}$. With this BH mass and the AGN bolometric luminosity derived earlier ($L_{\rm bol, AGN} = 5.5\times 10^{42}\, \rm erg s^{-1}$), we compute an Eddington ratio within the range of $\rm \log_{10} \lambda_{Edd}  \approx (-2.83,-2.36)$, consistent with the system's previously accepted status as a relatively weak AGN. This sets ESO 428-G14 in a region of the $\rm N_{H}-\lambda_{Edd}$ plane where outflows are not efficiently driven.

It is worth noting that the substantial systematic uncertainties on M$_{\rm BH}$, $L_{\rm bol, AGN}$ and N$_{\rm H}$ can shift the position of the AGN around on this diagram. The strength of the wind and its orientation depend on $\rm \lambda_{Edd}$ and $\rm n_{H}$ \citep[e.g.][]{Ricci+17,Venanzi+20}, which are not well-constrained for the heavily-obscured AGN in ESO 428-G14. We also note that, while the $\rm N_{H}-\lambda_{Edd}$ plane defines the ideal region for launching radiation driven winds, such outflows are still found in AGN that lie outside the region. \\

 Torus models that include a self-consistent wind component predict the wind originates at the sublimation edge of the torus \citep{Honig+13,Honig+17}. 
Consequently, the composition of the dust within the wind is expected to be similar to that of the sublimation region, where small dust grains, particularly silicates, are easily destroyed \citep[e.g.][]{Honig+17,2023A&A...676A..73G}. 
As such, it is primarily the larger dust grains of  size ($a \approx 0.075-1\,\rm  \mu m$) and graphites that are robust enough to survive in the sublimation region and as a result, in the dusty wind \citep[e.g.][]{Honig+17,Honig+19}.
Using Eqs.~1 \& 2, but taking the larger grain sizes expected in the wind, the temperature of the polar dust would be $T_{\rm dust} =[44,26]\, \rm K $ at $D \approx 100\, \rm pc$ from the nucleus. This is much lower than the estimated temperature of the extended polar dust in NW2. Therefore, we do not expect that the dust we see in the extended MIR emission is deficient in small grains. This adds credence to its origin as ambient ISM dust.
While a polar dusty wind may still be present in ESO 428-G14, it would be too small for our images to resolve,
lying well within the nuclear aperture as defined in Section \ref{regions}.

\subsection{Dust heated by shocks}
Fig.~\ref{dustonly-image} reveals the polar dust to be co-spatial with the radio jet, which could indicate that shocks induced by jet-ISM interactions are an important culprit in heating the dust.  
Indeed, previous studies have shown that the radio jet in ESO 428-G14 is coupled with the disk and shows spatial correspondence with a number of emission lines, indicating that the radio jet can strongly perturb the ionised gas \citep[e.g.][]{Falcke+96,Riffel+06,May+18}. \\

Evidence of shocks in ESO 428-G14 in the central $\rm \sim 200\,pc$ has been reported in a number of previous studies.
For instance, \citet{Riffel+06} find that the radio jet can contribute up to 90\% to the excitation of [\ion{Fe}{II}]1.257~$\mu$m, with strong enhancement in the [\ion{Fe}{II}]/Pa$\beta$ ratio along the jet region. Additionally, CLOUDY modelling by \citet{May+18} predict a strong dependence on shocks, where AGN photoionisation models alone fail to reproduce the observed [\ion{Si}{VI}]/Br$\gamma$ flux ratios at the 100 pc scale. Moreover, spectral analysis of the nuclear and extended X-ray emission in \citet{Fabbiano+18a} also predict shocks to be present within the central 170 pc region of ESO 428-G14 with velocities of at least a few hundred km~s$^{-1}$. In fact, \citet{Riffel+06} report on the detection of outflows towards the North-West with velocities up to  400 km~s$^{-1}$ which they link back to the radio hot spot. This is also supported by the detection of mechanically-driven [\ion{Si}{VI}] outflows within the boundaries of the radio emission, with velocities up to 250 km~s$^{-1}$ in the North-West and 100 km~s$^{-1}$ in the South-East \citep{May+18}.  \\

All of the above supports the possibility that the polar dust seen in the F1000W image can be driven by jet-ISM interactions. Moreover, it has been established that shocks can be a significant source of high-energy radiation \citep[e.g.][]{Dopita95,DopitaSutherland95}.

Following \citet{DopitaSutherland95}, the total radiative output per unit area of a shock ($\rm F_{T}$) is given by:
\begin{equation}
   F_T = 2.28 \times 10^{-3} \left( \frac{v_s}{100 \, \text{km s}^{-1}} \right)^{3.0} \left( \frac{n}{\text{cm}^{-3}} \right) \, \text{ergs cm}^{-2} \text{s}^{-1},
\end{equation}
where $v_s$ is the velocity of the shock and $n$ is the particle number density. An important assumption here is that the post-shock gas is strongly radiative, radiating away all its thermalised energy efficiently. \\

Given that previous studies report high ionised gas outflow velocities associated with the NW region ($\rm \sim 100 \, pc$) from the nucleus, we adopt measurements from \citet{May+18} (their  Table 2, region b1) for $v_s= 250\,\rm  km~s^{-1}$  and $n = 4350\,\rm  cm^{-3}$, where we have assumed that the highest ionised gas velocities are representative of the shock velocity, and that gas particle densities are equal to the electron densities measured from low ionisation gas tracers. We then compute $F_T = 156\, \rm erg/s/cm^{2}$. 
Assuming that the shocks are uniformly distributed through the region, so that their radiation can be treated as a uniform field, we can express $F_{\rm T}$ in terms of the Habing field, so that $G_{0} = 653\, F_{T}$.
Using Eq. 2, and taking a normal ISM grain size distribution \citep[a = 0.005-0.25 $\mu$m][]{DraineLee1984},
the expected temperature of the dust driven by shock heating  is $T_{\rm dust} = [154,70]\,\rm K$. 

The radiation environment arising from fast shocks in dense gas in NW2 is expected to be significantly higher than that arising from the AGN, following the reasonable values and assumptions taken in our calculations. If so, the temperatures of dust we measure in the extended emission may indeed only be possible due to the enhanced flux of higher energy photons arising from local shocks from the jet-driven outflow. \\

The strong enhancement of the MIR emission along the jet has been reported in other nearby AGN, notably NGC~1068 \citep{2000A&A...363..926A} and Cygnus~A \citep{1990MNRAS.246..163T}.
At face value, this association of dust emission with outflows is a bit surprising, because shock models predict that dust should be destroyed by the passage of the radio jet in the high temperatures of the post-shock gas  \citep[e.g.][]{1997ApJS..110..287F,2001MNRAS.328..848V}.
However, observations show little solid evidence of dust destruction. A few possible reasons for the empirical existence of warm dust emission around radio jets were put forward by \citet{2001MNRAS.328..848V}. If the dust is confined to dense cores of gas which are resistant to the full brunt of shocks, they may survive the passage of the jet. Alternatively, the timescale for dust destruction may be longer than the passage of the jet, or dust reformation could be efficient in the post-shock material. We also propose an additional possibility. Much of the line emission around radiative shocks is generated in gas that is ionised and illuminated by the shocks, but not directly within the post-shock gas \citep[the so-called ``precursor'' as defined by][]{DopitaSutherland95}. The dust mixed into this gas would not be destroyed by collisional processes, and can still reprocess the radiation from the shocks effectively. Therefore, if post-shock gas only occupies a modest filling factor in a jet-dominated region of the NLR, such as NW2, the dust emission we see could arise in the precursor. \\

\begin{table}
\centering
\caption{The temperatures derived for the different heating models and dust grain sizes considered.}
\begin{tabular}{l c c}
\toprule
Dust Source & Grain Size ($\mu$m) & Temperature (K) \\
\midrule
Illuminated in NLR   & 0.005, 0.2  & 76, 35  \\
From the torus       & 0.075, 1    & 44, 26  \\
Heated by shocks     & 0.005, 0.2 & 152, 70 \\
\bottomrule
\label{table:temp}
\end{tabular}
\end{table}

Additional evidence for the role of local heating from shocks is the fact that, despite strong coronal [\ion{Si}{VI}] emission towards the South-East NLR in ESO 428-G14, the dust continuum emission in that region is weak. The coronal lines can only be produced by the hard UV field of the AGN. Yet, despite clear AGN-ionised gas in that region, there is little warm dust. The region to the North-West, in contrast, has both coronal line emission and warm dust emission. This asymmetry is mirrored by the structure of the radio jet, which is brighter and well-collimated only towards the North-West. We can reconcile this empirical difference by positing that a major part of the dust emission is heated directly by energy connected to the jet, such as through shocks. 

If confirmed, our JWST observations offer a unique insight into the feedback efficiency of the jet, in both this AGN and potentially other systems that may show similar properties.\\

To summarise, if we assume typical ISM grains with a sizable fraction of small grains, then local shock heating is more likely to achieve the temperature of $\sim 120 \,\rm K$ that we find in the North-Western extended dust emission, at distances of $100\, \rm pc$ from the nucleus (see Table~\ref{table:temp}). However, this is under the idealised assumption that all the thermal energy produced in jet-driven shocks is radiated away \citep{DopitaSutherland95}, and reprocessed by dust in the vicinity of the shocks. Some of the post-shock radiation will escape the NLR, while, in turn, X-ray and hard UV heating of dust by the AGN's radiation may produce a substantial tail of warm grains. As such, heating from the central engine is also expected to contribute to the observed dust emission.

\section{Conclusion}
\label{sec:5}

ESO 428-G14 shows a complex and unique NLR defined by its ``braided'' strands and  radio jet. In this work, we have investigated the morphology of the dust within the NLR, focusing on the inner $4 \arcsec \times 4 \arcsec$ region, to provide imaging evidence of polar MIR emission, reveal its nature, and explore its origin. Our study combines JWST/MIRI data from the GATOS survey, using multi-band imaging in five filters (F560W, F1000W, F1500W, F1800W, F2100W), with ancillary data from  several facilities (ALMA, HST, SINFONI, VLA,  Spitzer) to contextualise the new MIR findings within the broader framework. Our main findings can be summarised as follows:

\begin{itemize}

\item We detect in the inner $4 \arcsec \times 4 \arcsec$ region extended MIR emission in all bands, with the exception of F560W (see top panel in Fig. ~\ref{SED_plot}). The new MIR emission reveals an asymmetric and collimated morphology, extending along a  PA of $\sim 131\degree$ from the North-West to the South-East, and  with a total extent of  $\sim \rm 450 $ pc (Fig. ~\ref{original_rgb}, and Fig.~\ref{SED_plot}). 

\item We find this MIR structure to be  polar, evidenced by its alignment with the radio jet and other ionised gas emission lines that define the NLR. Comparison with the ALMA CO(2-1) nuclear emission,  which is a proxy for the torus orientation, reveals the MIR structure to be perpendicular to it, reinforcing its polar orientation (see Fig.~\ref{jwst_plus_anci}).

\item We derive photometric SEDs across various regions of interest and observe that regions along the polar structure  are characterised by SEDs consistent with cooler dust relative to that in disk regions. Assuming a simple black body, we estimate temperatures of the polar structure to be $\sim 100-120\, \rm K$. (see regions SE1 and NW2 in Fig.~\ref{SED_plot}).

\item  We investigate emission line contamination in the F1000W image and find that the extended emission ($\sim 1''$ from the nucleus) towards both the SE (see SE2, SE3 in Fig.~\ref{SED_plot}) and NW (see NW1, NW2 in Fig.~\ref{SED_plot}) regions is contaminated by  [\ion{S}{IV}], reaching a level of the order of $\sim 50\%$ (see Fig.~\ref{contamination_map}).

\item We generate a pure dust continuum image from the F1000W filter and uncover that the dust is more compact, extending up to $\sim 1''$ on each side of the nucleus. This polar dust appears to be co-spatial with the radio jet (see Fig. ~\ref{dustonly-image}).

\item We explore the possible heating processes that can contribute to the polar dust reaching temperatures of $\sim 120\, \rm K$ at $D= 100\, \rm pc$ from the nucleus and predict, under the assumption of typical ISM grain size, that local heating by jet-induced shocks can reproduce similar temperatures (see Table~\ref{table:temp}). Another heating process, most likely arising from illumination through AGN radiation fields, may also be at play.
\end{itemize}

Future work on self-consistent modelling of radiative transfer through dusty NLR gas with different grain size distributions, fine tuned to the nuclear system in ESO 428-G14, will allow us to understand the origin and
main heating mechanism of the extended MIR dust emission in this AGN.
Moreover, while our findings indicate significant emission line contamination in ESO 428-G14, it is  crucial to recognise that such contamination may not be a universal feature in all JWST/MIRI images. Indeed, the extent of contamination could very well be object dependent, varying depending on its specific AGN characteristics and geometry. Upcoming  GATOS sample studies  are poised to offer a comprehensive study of contamination.

\section{DATA AVAILABILITY}
All data used in this paper can be extracted as described in section ~\ref{data} and ~\ref{sec:archival-data}. The reduced JWST/MIRI images are available online on Zenodo under the Creative Commons Attribution 4.0 International
license, DOI: 10.5281/zenodo.11491161.

\section{Acknowledgments}
We are grateful to the referee for their valuable suggestions that have significantly improved this manuscript. HH would also like to thank Tiago Costa for insightful discussions on shock heating and Devang H. Liya for helpful feedback. 
HH acknowledges support and funding from STFC and Newcastle University.
HH, DJR \& SC acknowledge the support of the UK STFC through grant ST/X001105/1.
AAH acknowledges
support from grant PID2021-124665NB-I00 funded by the Spanish
Ministry of Science and Innovation and the State Agency of Research
MCIN/AEI/10.13039/501100011033  and ERDF A way of making Europe.
MPS acknowledges support from grant RYC2021-033094-I funded by MICIU/AEI/10.13039/501100011033 and the European Union NextGenerationEU/PRTR.
IGB acknowledges support from STFC through grants ST/S000488/1 and ST/W000903/1.
CR acknowledges support from Fondecyt Regular grant 1230345 and ANID BASAL project FB210003.
MS acknowledges support by the Ministry of Science, Technological Development and Innovation of the Republic of Serbia (MSTDIRS) through contract no. 451-03-66/2024-03/200002 with the Astronomical Observatory (Belgrade).
AJB acknowledges funding from the ``FirstGalaxies'' Advanced Grant from the European Research Council (ERC) under the European Union’s Horizon 2020 research and innovation program (Grant agreement No. 789056).
DR acknowledges support from STFC through grants ST/S000488/1 and ST/W000903/1.
C.P., L.Z., and M.T.L  acknowledge grant support from the Space Telescope Science Institute (ID: JWST-GO-02064.002).
SGB acknowledges support from the Spanish grant PID2022-138560NB-I00,
funded by MCIN/AEI/10.13039/501100011033/FEDER, EU. 
C.M.H acknowledges funding from a United Kingdom Research and Innovation grant (code: MR/V022830/1).
OG-M acknowledge financial support from PAPIIT UNAM project IN109123 and “Ciencia de Frontera” CONAHCyT project CF-2023-G-100. 
ARA acknowledges Conselho Nacional de Desenvolvimento Científico e Tecnológico (CNPq) for partial support to this work through grant 313739/2023-4.
E.B. acknowledges the Mar\'ia Zambrano program of the Spanish Ministerio de Universidades funded by the Next Generation European Union and is also partly supported by grant RTI2018-096188-B-I00 funded by the Spanish Ministry of Science and Innovation/State Agency of Research MCIN/AEI/10.13039/501100011033.
This work is based [in part] on observations made with the NASA/ESA/CSA James Webb Space Telescope.
This paper makes use of the following ALMA data: ADS/JAO.ALMA\#2015.1.00086.S. ALMA is a partnership of ESO (representing its member states), NSF (USA) and NINS (Japan), together with NRC (Canada), MOST and ASIAA (Taiwan), and KASI (Republic of Korea), in cooperation with the Republic of Chile. The Joint ALMA Observatory is operated by ESO, AUI/NRAO and NAOJ. 
This work makes use of several \texttt{PYTHON} packages:  \texttt{NumPy} \citep{Harris2020}, \texttt{Matplotlib} \citep{Hunter2007}, and \texttt{Astropy} \citep{Astropy2013,Astropy2018}. \\

\bibliography{references}


\hfill \\

\noindent
$^1$ School of Mathematics, Statistics and Physics, Newcastle University, Newcastle upon Tyne, NE1 7RU, UK\\
$^2$ Centro de Astrobiolog\'{\i}a (CAB), CSIC-INTA, Camino Bajo del
Castillo s/n, E-28692 Villanueva de la Ca\~nada, Madrid, Spain \\
$^{3}$ Instituto de F\'isica Fundamental, CSIC, Calle Serrano 123, 28006 Madrid, Spain  \\
$^4$ Department of Physics, University of Oxford, Keble Road, Oxford, OX1 3RH, UK \\
$^{5}$ Department of Physics \& Astronomy, University of Southampton, Highfield, Southampton SO171BJ, UK \\
$^{6}$ Instituto de Astrof\'isica de Canarias, Calle V\'ia L\'actea, s/n, E-38205, La Laguna, Tenerife, Spain \\
$^{7}$ Departamento de Astrof\'isica, Universidad de La Laguna, E-38206 La Laguna, Tenerife, Spain \\
$^8$ Department of Physics \& Astronomy, University of Alaska Anchorage, Anchorage, AK 99508-4664, USA \\
$^9$ Department of Physics, University of Alaska, Fairbanks, Alaska 99775-5920, USA \\
$^{10}$ Department of Physics and Astronomy, The University of Texas at San Antonio, 1 UTSA Circle, San Antonio, Texas, 78249-0600, USA\\
$^{11}$ Max Planck Institute for Extraterrestrial Physics (MPE), Giessenbachstr.1, 85748 Garching, Germany\\
$^{12}$ Instituto de Estudios Astrof\'isicos, Facultad de Ingenier\'ia y Ciencias, Universidad Diego Portales, Av. Ej\'ercito Libertador 441, Santiago, Chile \\
$^{13}$ Kavli Institute for Astronomy and Astrophysics, Peking University, Beijing 100871, China \\
$^{14}$ Kavli Institute for Particle Astrophysics \& Cosmology (KIPAC), Stanford University, Stanford, CA 94305, USA \\
$^{15}$ Observatorio Astronómico Nacional (OAN-IGN)-Observatorio de  Madrid, Alfonso XII, 3, 28014-Madrid, Spain \\
$^{16}$ National Astronomical Observatory of Japan, National Institutes of Natural Sciences (NINS), 2-21-1 Osawa, Mitaka, Tokyo 181-8588, Japan\\
$^{17}$ Departmento de F\'isica de la Tierra y Astrof\'isica, Fac. de CC F\'isicas, Universidad Complutense de Madrid, E-28040 Madrid, Spain\\
$^{18}$ Instituto de F\'isica de Partículas y del Cosmos IPARCOS, Fac. CC F\'isicas, Universidad Complutense de Madrid, E-28040 Madrid, Spain\\
$^{19}$ Cahill Center for Astrophysics, California Institute of Technology, 1216 East California Boulevard, Pasadena, CA 91125, USA \\
$^{20}$ Astronomical Institute, Academy of Sciences, Bo\v{c}n\'{i} II 1401, CZ-14131 Prague, Czech Republic \\
$^{21}$ LERMA, Observatoire de Paris, Coll\`ege de France, PSL
University,  CNRS, Sorbonne University, Paris \\
$^{22}$ Institute of Astrophysics, Foundation for Research and Technology-Hellas (FORTH), Heraklion, GR-70013, Greece \\
$^{23}$ School of Sciences, European University Cyprus, Diogenes street, Engomi, 1516 Nicosia, Cyprus \\
$^{24}$ Instituto de Radioastronom\'ia y Astrof\'isica (IRyA), Universidad Nacional Aut\'onoma de M\'exico, Antigua Carretera a P\'atzcuaro \#8701, Colonia ExHda. San Jos\'e de la Huerta, Morelia, Michoac\'an, M\'exico C.P. 58089 \\
$^{25}$ Telespazio UK for ESA, ESAC, Camino Bajo del Castillo s/n, 
28692 Villanueva de la Cañada, Spain.\\
$^{26}$ Space Telescope Science Institute, 3700 San Martin Drive, Baltimore, MD 21218, USA \\
$^{27}$ Gemini Observatory/NSF’s NOIRLab, Casilla 603, La Serena, Chile \\
$^{28}$ School of Sciences, European University Cyprus, Diogenes street, Engomi, 1516 Nicosia, Cyprus\\
$^{29}$ Laborat\'{o}rio Nacional de Astrof\'{i}sica/MCTI, 37530-000, Itajub\'{a}, MG, Brazil \\
$^{30}$ Observat\'{o}rio Nacional, Rua General Jos\'{e} Cristino, 77, 20921-400 S\~{a}o Crist\'{o}v\~{a}o, Rio de Janeiro, RJ, Brazil. \\
$^{31}$ Astronomical Observatory, Volgina 7, 11060 Belgrade, Serbia \\
$^{32}$ Sterrenkundig Observatorium, Universiteit Gent, Krijgslaan 281-S9, Gent B-9000, Belgium\\
$^{33}$ Centre for Extragalactic Astronomy, Department of Physics, Durham University, South Road, Durham, DH1 3LE, UK\\

\appendix

\section{Saturation}
\noindent

\begin{figure*}
\label{saturation}
  \centering
  {\includegraphics[width=0.98\textwidth]{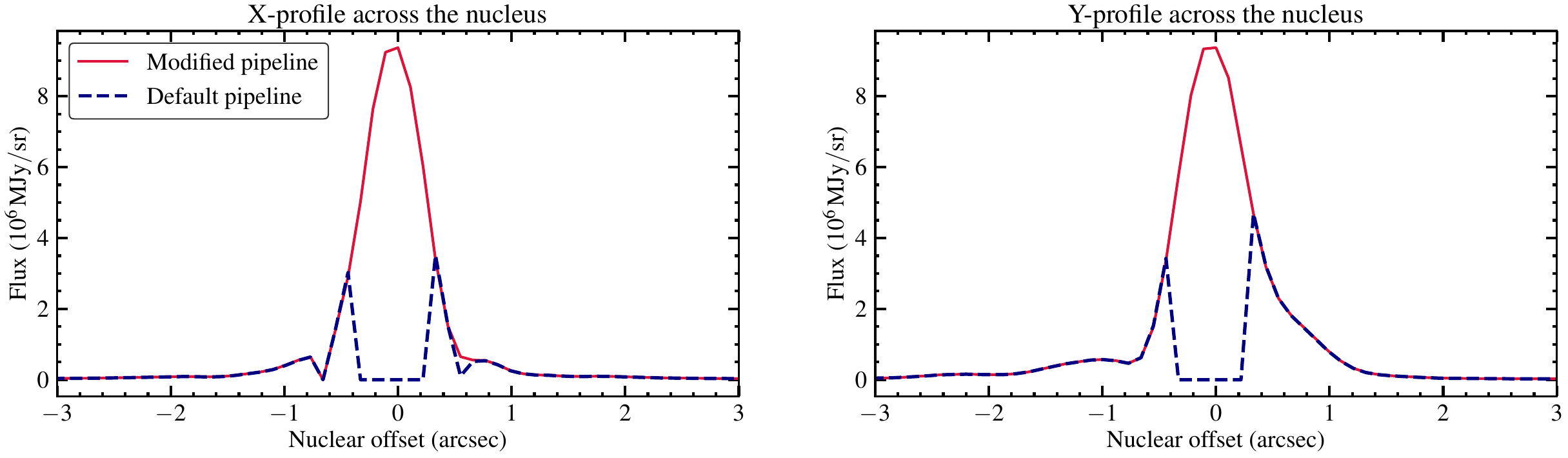}}
-
  \caption{ X- and Y- profiles (left and right panels) through the nucleus of ESO 428-G14 for the F15000W filter using the standard JWST pipeline, where the output image is saturated in the nucleus (see blue dashed line) compared to the modified pipeline where first frame corrections are turned off (see red line). This modified approach was adopted to gain at least two good groups for the linear ramp fitting, thereby ensuring an estimate of the flux despite the presence of other saturated groups.}  
  \label{saturation}
\end{figure*}
The JWST pipeline consists of three main stages. Stage\,1 performs detector level-corrections, stage 2 applies flux calibrations, and finally, stage 3 combines all frames together to form one image. Our observations consist of four frames, each frame consists of 50 integrations, and each integration consists of 5 groups. By default, the first (group 1) and last group (group 5) are flagged as ``do not use'' by the  pipeline to avoid transient effects between exposures. As such, if  saturation starts at group 3, then also groups 4 and 5 are flagged as saturated. In this case, only group 2 is flagged as good, and the pipeline fails to fit a straight line with only one point. To provide the ramp fitting with at least two good groups, we turn off the `first frame" correction which ordinarily flags the first group. This way, we have at least 2 groups (group 1 and 2) to perform a linear fit, allowing us to retrieve flux measurements from these otherwise compromised pixels. While using only two groups for the ramp fitting may introduce some uncertainties in the linearity correction step due to the limited data points, this approach still provides valuable flux estimates from otherwise unusable saturated pixels.  \\

Fig. ~\ref{saturation} shows the X and Y intensity profiles across the nucleus in the F1500W image of ESO 428-G14, which experienced the most saturation in the nucleus. The default JWST pipeline results, depicted with dashed blue lines, show where this saturation occurs. We modify  the pipeline by turning off the first frame correction for these pixels, thereby allowing us to gain more valid groups for the ramp fitting process (see red line). This modification ensures that at least two groups are available for a reliable fit.

\section{DISK BLOBS}
\begin{figure*}
  \centering
  {\includegraphics[width=0.6\textwidth]{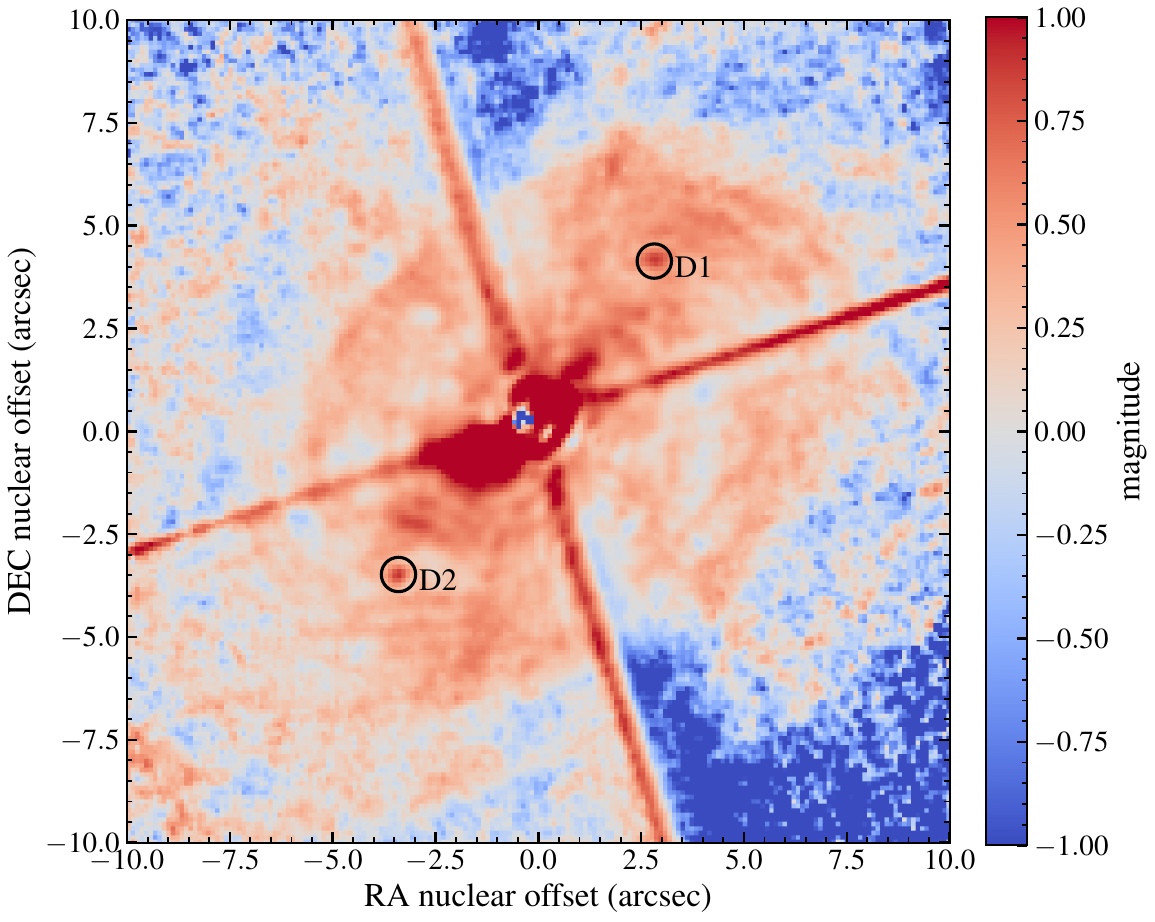}}
  \caption{F560W-F1000W colour image of ESO 428-G14 expressed in units of magnitude. The circles mark the locations of D1 and D2, also identified in Fig.~\ref{blob-SB}. }
  \label{f560W-f1000W-ratio}
\end{figure*}

Fig. ~\ref{f560W-f1000W-ratio} shows a colour image using the filters F560W and F1000W. The flux values from both images are converted to magnitudes using $\rm M = - 2.5\, \log_{10}(F)$, where F is flux. The image is colour coded such that red indicates redder (cooler) magnitudes, while blue indicates bluer (hotter) temperatures. This highlights the SF disk regions D1 (in the North-West) and D2 (in the South-East) presented in Fig.~\ref{blob-SB}, making them stand out against the disk. Interestingly, D1 and D2 are along a similar PA to the extended polar dust. This alignment hints to the contribution of AGN activity in their heating. Recent paper by \citet{HermosaMunoz2024} reports on the detection of SF knots along the direction of the outflow in NGC 7172. Future kinematic analysis of ESO 428-G14 would allow us to shed light onto the nature of these regions.

\end{document}